\documentclass{aa}
\usepackage{graphicx}
\usepackage{txfonts}
\usepackage{natbib}
\usepackage{url}
\usepackage{aalongtable}
\usepackage{rotating}
\usepackage{hyperref}

\begin{document}

   \title{Tracing the evolution of NGC6397 through the chemical
     composition of its stellar populations \thanks{Based on data
       collected at European Southern Observatory (ESO), Paranal,
       Chile, under program IDs 077.A-0018(A) and 281.D-5028(A), as
       well as data collected with the Danish 1.54\,m at European
       Southern Observatory (ESO), La Silla.}\fnmsep\thanks{Two tables
       with line equivalent widths, chemical abundances, and stellar
       parameters are only available in electronic form
       at the CDS via anonymous ftp to cdsarc.u-strasbg.fr (130.79.128.5)
       or via \url{http://cdsweb.u-strasbg.fr/cgi-bin/qcat?J/A+A/}}}

   \author{K. Lind\inst{1,6}\and
           C. Charbonnel\inst{2,3}\and
           T. Decressin\inst{4}\and
           F. Primas\inst{1}\and
           F. Grundahl\inst{5}\and
           M. Asplund\inst{6}
           }

   \institute{European Southern Observatory (ESO),
              Karl-Schwarzschild-Strasse 2,
              857 48 Garching bei M\"unchen, Germany\and
              Geneva Observatory, 51 chemin des Mailettes, 1290
              Sauverny, Switzerland\and
              Laboratoire d'Astrophysique de Toulouse-Tarbes, CNRS UMR
              5572, Universit\'e de Toulouse, 14, Av. E.Belin, F-31400
              Toulouse, France\and
              Argenlader Institut f\"ur Astronomie (AIfA),
              Universit\"at Bonn, Auf dem H\"ugel 71, 531\,21 Bonn,
              Germany\and 
              Department of Physics \& Astronomy, Aarhus University,
              Ny Munkegade, 8000 \AA rhus C, Denmark\and
              Max-Planck-Institut f\"ur Astrophysik,
              Karl-Schwarzschild-Strasse 1,
              857 41 Garching bei M\"unchen, Germany\\
              \email{klind@mpa-garching.mpg.de}
             }
 
   \date{Received 07/07/2010, accepted 28/11/2010}

\authorrunning{K.\,Lind et al.}  \titlerunning{The chemical evolution of NGC\,6397.}

  \abstract
   {The chemical compositions of globular clusters provide important
     information on the star formation that occurred at very early
     times in the Galaxy. In particular the abundance patterns of
     elements with atomic number $\rm z\le13$ may shed light on the
     properties of stars that early on enriched parts of the
     star-forming gas with the rest-products of hydrogen-burning at
     high temperatures.}  {We analyse and discuss the chemical
     compositions of a large number of elements in 21 red giant branch
     stars in the metal-poor globular cluster NGC\,6397. We compare
     the derived abundance patterns with theoretical predictions in
     the framework of the ``wind of fast rotating massive
     star"-scenario.}  {High-resolution spectra were obtained with the
     FLAMES/UVES spectrograph on the VLT. We determined non-LTE
     abundances of Na, and LTE abundances for the remaining 21
     elements, including O (from the [OI] line at 630\,nm), Mg, Al,
     $\alpha$, iron-peak, and neutron-capture elements, many of which
     had not been previously analysed for this cluster. We also
     considered the influence of possible He enrichment in the
     analysis of stellar spectra.}  {We find that the Na abundances of
     evolved, as well as unevolved, stars in NGC\,6397 show a distinct
     bimodality, which is indicative of two stellar populations: one
     primordial stellar generation of composition similar to field
     stars, and a second generation that is polluted with material
     processed during hydrogen-burning, i.e., enriched in Na and Al
     and depleted in O and Mg. The red giant branch exhibits a similar
     bimodal distribution in the Str\"{o}mgren colour index
     $c_y=c_1-(b-y)$, implying that there are also large differences
     in the N abundance. The two populations have the same composition
     for all analysed elements heavier than Al, within the measurement
     uncertainty of the analysis, with the possible exception of
     [Y/Fe]. Using two stars with almost identical stellar parameters,
     one from each generation, we estimate the difference in He
     content, $\Delta Y=0.01\pm0.06$, given the assumption that the
     mass fraction of iron is the same for the stars.}  {NGC\,6397
     hosts two stellar populations that have different chemical
     compositions of N, O, Na, Mg, and probably Al. The cluster is
     dominated (75\%) by the second generation. We show that massive
     stars of the first generation can be held responsible for the
     abundance patterns observed in the second generation long-lived
     stars of NGC\,6397. We estimate that the initial mass of this
     globular cluster is at least ten times higher than its
     present-day value.}


   \maketitle

\section{Introduction}
\label{sec:intro}
The abundance patterns of light elements (up to Al) in globular
clusters are in the process of being carefully investigated. In
particular, many groups are studying the origin of the larger spread
in C, N, O, Na, Mg, and Al abundances compared to field stars of
similar metallicity. The present status of the observed light element
(O, Na, Mg, and Al) abundances in globular clusters and their possible
consequences for the formation and enrichment history of these stellar
populations have been presented in a series of publications by
\citet[e.g.][]{Carretta09b,Carretta10b} (see also reviews by
\citealt{Gratton04} and \citealt{Charbonnel05b}). The main findings,
as inferred from high-resolution spectroscopy of individual globular
cluster stars, are apparent enhancements of N, Na, and Al abundances
and deficiencies in Li, O, and Mg. These patterns can be naturally
explained in terms of the enrichment of the nucleosynthesis
rest-products of H-burning at high temperatures
\citep{Denisenkov89,Langer93}. Since the resulting anti-correlations
between the O--Na abundances in particular have not only been seen in
evolved red giant branch (RGB) stars, but also in turn-off (TO) and
subgiant branch (SGB) stars \citep{Gratton01}, an intrinsic stellar
evolutionary cause is very unlikely, i.e. the stars cannot have
established these abundance patterns themselves. It is instead
believed that the gas that formed second generation stars in globular
clusters underwent early pollution by slow ejecta from intermediate or
massive stars.

In this context, it is significant that photometric evidence of
multiplicity has been found in some clusters, e.g. the parallel main
sequences identified in $\omega$ Cen and NGC\,2808. These
observations seemingly necessitate that a difference in the He content
be present \citep[e.g.][]{Norris04}, in qualitative agreement with the
self-enrichment process responsible for the other light-element
variations. Helium enrichment is also commonly invoked to explain the
extended horizontal branch observed in many clusters
\citep{DAntona08}. We are, however, far from building a fully
consistent picture of the chemical evolution of globular clusters that
can explain all the various observations simultaneously.

A key unknown is the nature of the polluting objects. One possibility
is that so-called hot bottom burning occurs at the base of the
convective envelope in intermediate-mass stars during the asymptotic
giant branch (AGB) phase, leaving nucleosynthesis products in the
envelopes that are subsequently expelled
\citep{Ventura08,Ventura10}. In addition, super-AGB stars have been
suggested to be responsible for the most extreme anomalies
\citep{Pumo08}. The main alternative scenario is a slow mechanical
wind from rapidly rotating massive stars \citep[e.g.][]{Decressin07b},
whose envelopes have been enriched in H-burning products by means of
deep internal mixing (see Sect.\,\ref{sec:conseq}). Yet another option
is mass loss from massive binary systems, as suggested by
\citet{deMink09}.

By performing accurate abundance analysis of many elements (and
isotopic ratios) in large stellar samples, we may be able to pin-point
the nature of the progenitors \citep[e.g.][and references
  therein]{Charbonnel05b}. A common property of the competing
scenarios is that the pollution would mainly alter the light element
abundances of the second generation stars in globular clusters, thus
leaving $\alpha$ and iron-peak elements unaffected. This is necessary
to explain the homogeneous composition of these elements seen in most
clusters. Elements created in the s-process may be affected by AGB
pollution, suggesting that correlations exist between s-process and
light element anomalies. In NGC\,6752, an unexplained correlation was
indeed identified between Al abundances and Y, Zr, and Ba
\citep{Yong05}, but the systematic heavy element variations are small
(0.1\,dex) and comparable to the statistical scatter. One must also
bear in mind that the yields of AGB stars are uncertain
\citep[e.g.][]{Charbonnel07b,Decressin09,Ventura10}.

In addition to mapping the presence of abundance trends and
correlations, it is also essential to investigate, preferably with
sound number statistics, the fraction of stars with normal chemical
compositions, similar to those in the field, and the fraction of
second and possibly third generation stars in globular
clusters. Linking this information to other cluster observables, one
may construct a schematic model for the episodes of star formation and
evolution. \citet{Carretta10b} describe a possible general formation
scenario in which a precursor population, forming from the gas
assembled at a very early epoch inside a CDM halo, efficiently raises
the metal content of the gas cloud via core-collapse supernova
explosions. These trigger a second, large episode of star formation,
the so-called primordial population (first generation). The slow winds
of massive or intermediate-mass stars of the primordial population
feed a cooling flow, and the intermediate (second) generation of stars
are formed in the central parts of the cluster, out of material
enriched in H-burning products. The remaining gas is dispersed by
core-collapse supernovae of the second generation and star formation
ceases. The present-day cluster is dominated by the second generation,
with a smaller fraction, approximately 30\%, being left of the
primordial population. Critical factors determining the outcome of
this scenario are the initial mass function (IMF) of the polluting
stars, the initial total mass of the cluster, and the amount of mixing
between processed gas in the slow stellar ejecta and pristine cluster
gas. We discuss these issues in Sect.\,\ref{sec:conseq}.

HST photometry of NGC\,6397 produces a remarkably clean HR-diagram,
with a very tight main sequence \citep{Richer08} and a very compact
blue horizontal branch, i.e. there are no obvious photometric signs of
multiple populations. The cluster is well-studied in terms of numbers
of stars for which high-resolution spectra have been obtained, but
only a handful of elements have previously been analysed even on the
RGB \citep[most recently
  by][]{Castilho00,Gratton01,Thevenin01,Korn07,Carretta09b}, as
summarised in Table \ref{tab:mabund}.

Early studies of the strengths of the G and CN band in RGB stars in
NGC\,6397 (and other clusters) suggested that there are anomalies in
their C and N abundances \citep{Bell79,Briley90}. Eventually,
\citet{Gratton01} also detected a significant spread in Na abundance
for a sample of ten TO stars and RGB stars, findings that clearly
pointed to an intrinsic, rather than evolutionary,
origin. \citet{Carretta05} corroborated these findings for a larger
sample of stars and also found a significant O--Na anti-correlation,
as well as a large spread in C and N abundance \citep[see
  also][]{Pasquini04}. In the latest analysis by
\citet{Carretta09a,Carretta09b}, the O--Na anti-correlation is
present, although the number statistics are still rather small, oxygen
measurements in particular being few in number. A Mg--Al
anti-correlation has not been identified in NGC\,6397, but Mg also
seems to exhibit a certain scatter \citep{Korn07}. \citet{Lind09b},
hereafter Paper I, presented Na abundances for $>100$ TO, SGB, and RGB
stars, and found that the most heavily Na-enriched stars are also
significantly depleted in Li. It is thus clear that NGC\,6397, like
other globular clusters, should no longer be regarded as a single
stellar population despite the tightness of its colour-magnitude
diagram. However, even if pollution indeed seems to have taken place
in the cluster, it is unclear to which extent, and which elements are
affected by it. In this study, which targets red giants, we cover as
many elements as possible for a large sample of stars and give a more
decisive answer to this question.

\section{Observations and analysis}
\label{sec:observ}
We obtained high-resolution ($R=47000$) spectra with the multi-object
fibre spectrograph FLAMES/UVES on the VLT, covering the wavelength
range 480--680\,nm (UVES Red 580 standard setting). For a subset of
six stars, we also have spectra covering the wavelength range
760--1000\,nm (UVES Red 860 standard setting). Our targets are all RGB
stars, whose surface abundances should reflect the initial composition
of cluster gas, with the notable exceptions of Li and Be, which are
not studied here (Li abundances were published in Paper I).  In
particular, any internal abundance gradients created by atomic
diffusion \citep{Korn07,Lind08} during the main sequence have been
erased by the growth of the convective envelope.

\begin{figure}
        \centering \includegraphics[width=6.5cm,angle=90]{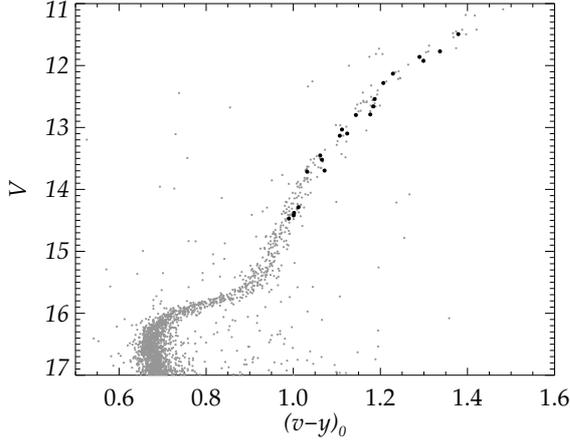}
          \caption{The targeted RGB stars are marked with black
            bullets in the $V-(v-y)$ diagram of NGC\,6397. The RGB
            bump is located at $V\approx12.6$.}
        \label{fig:y}  
\end{figure}

We consider here a subset of the targets analysed in Paper I, and we
refer to that study for a detailed description of the observations,
data reduction, and stellar parameter determination. In brief, we
determined effective temperatures from the $(b-y)$ photometric index,
interpolating between data points in the grid of synthetic
Str\"{o}mgren colours computed by \citet{Onehag09} using MARCS model
atmospheres \citep{Gustafsson08}. Surface gravities were derived from
the standard relation to mass and luminosity. In total, we present
abundances for 21 red giants distributed along the upper RGB of
NGC\,6397 (see Fig.\,\ref{fig:y}), above and below the bump. The
targets span a total range of approximately 500\,K in effective
temperature and 1.5\,dex in surface gravity.

The full list of lines used in the analysis is given in Appendix A
(except for all iron lines, which are listed in Paper I), with
references to the oscillator strengths that were adopted. The
wavelengths and the majority of $f$-values were taken from the Vienna
Atomic Line Database (VALD). In all abundance analyses we used 1D,
LTE, spherically symmetric, hydrostatic model atmospheres computed
with the MARCS code \citep{Gustafsson08}. For O, Na, Mg, and Al lines,
we applied spectrum synthesis with the Uppsala code \textsc{Bsyn},
whereas all other elemental abundances were determined from equivalent
width measurements of single lines that were translated to abundances
with the corresponding code \textsc{Eqwidth}. Abundance ratios with
respect to the Sun were calculated using the solar abundances of
\citet{Asplund09}. The listed abundance errors reflect only the
measurement uncertainty, as estimated from the signal-to-noise ratio
of the spectra \citep{Norris01}, and do not consider uncertainties in
the adopted stellar parameters. Upper limits to the equivalent width,
hence the abundance, were calculated by adding the estimated
measurement uncertainty twice to the best-fit synthetic spectrum.  Our
analysis has its obvious limitations in that it relies mostly on
traditional 1D modelling and LTE line formation, and neglects
hyperfine splitting. However, the modelling procedure is fully
adequate for establishing a robust internal abundance scale, which is
our primary objective.

\begin{figure}
        \centering \includegraphics[width=17cm,angle=90,viewport=0cm
          3cm 26cm 21cm]{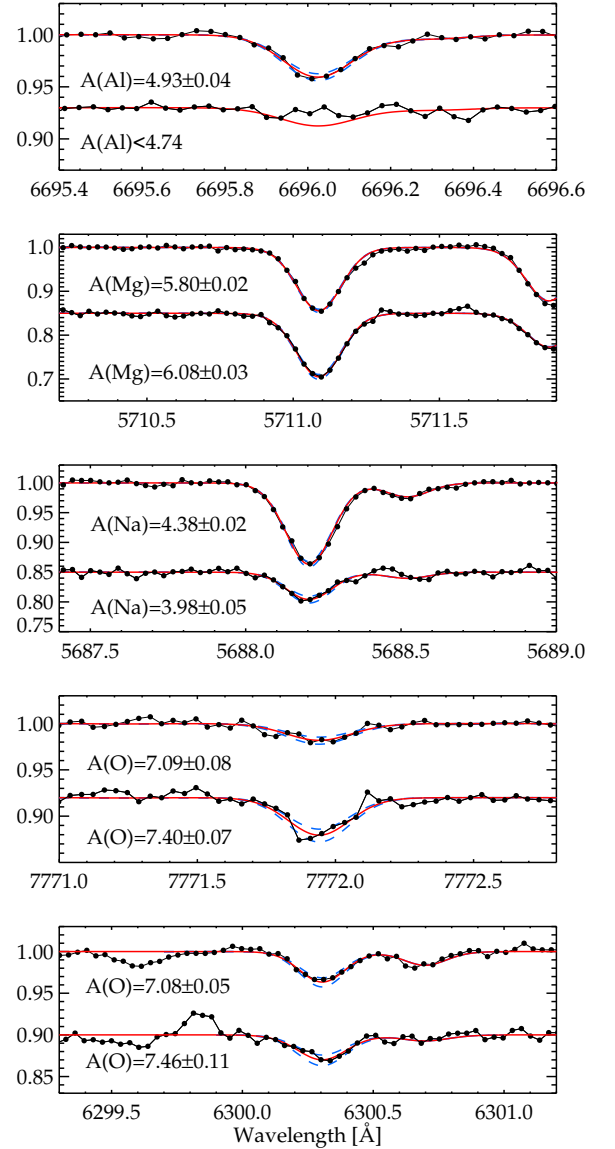}
          \caption{The figure shows some sample UVES spectra and the
            best fitting synthetic models and $\pm 1\sigma$
            profiles. The top spectrum in each panel represents a
            second generation star (12138) and the bottom a first
            generation star (17691).}
        \label{fig:specs}                                                                        
\end{figure}

Fig.\,\ref{fig:specs} shows the observed and best-fit model spectra of
O, Na, Mg, and Al lines in two RGB stars, \#17691 and \#12138. The
abundances of these four elements were all based on weak lines, whose
line formation do not depart substantially from LTE. For Na only, we
computed and adopted non-LTE abundances for all targets, whereas LTE
was assumed to hold for the other species. The oxygen abundance was
determined from the forbidden line at 630\,nm, which is believed to
have negligible non-LTE corrections, but may be sensitive to
granulation effects \citep{Kiselman01,Nissen02}. For the four targets
for which this was possible, we also determined the oxygen abundance
based on the NIR triplet lines at 777.1\,nm and 777.4\,nm, finding
very good agreement with the 630\,nm line in LTE. \citet{Fabbian09}
predict non-LTE corrections for the triplet lines to be of the order
of $-0.05...-0.08$\,dex for our targets, adopting hydrogen collisions
with $S_{\rm H}=1.0$ \citep{Pereira09b}. Applying this correction
would slightly lessen the agreement with the forbidden line. However,
a different choice of oscillator strength \citep[e.g. that
  by][]{Storey00}, would lower the oxygen abundances inferred from the
630\,nm line by a similar amount.

\begin{figure}
        \centering
                \includegraphics[width=6.5cm,angle=90]{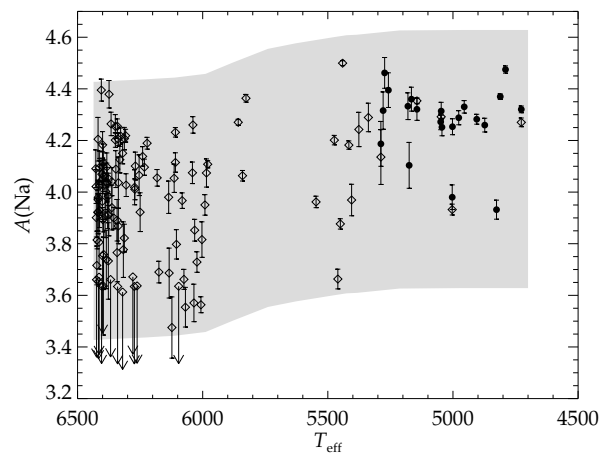}
          \caption{Non-LTE Na abundance variation with effective
            temperature in NGC\,6397. Bullets represent the RGB
            stars that are analysed in this study (abundance based on
            568.2/568.8\,nm lines) and open diamonds the stars
            analysed in Paper I (\citealt{Lind09b}, abundance based on
            818.3/819.4\,nm lines). The studies have five targets in
            common. The grey shaded area illustrates, given a range of
            different initial Na abundances, the predicted evolution
            of the surface abundance based on a model with atomic
            diffusion and turbulence below the outer convective zone
            \citep[][model T6.0]{Richard05a}}
        \label{fig:NaT}                                                                                       
\end{figure}

The Na abundances were based on the 568.2\,nm and 568.8\,nm doublet
lines by interpolating between non-LTE curve-of-growths, established
with the most recent atomic data for a grid of MARCS model atmospheres
\citep{Lind10b}. The non-LTE corrections are moderate, being of the
order of $-0.07$\,dex for stars with these stellar
parameters. For five targets, we compared the abundances inferred from
the 568.2/568.8\,nm lines with those inferred from the 818.3/819.4\,nm
lines in Paper I.  To ensure that the set of Na abundances are fully
consistent with the current study, we used the published equivalent
widths in Paper I and recomputed the abundances based on the latest
non-LTE calculations. For these near-infra red lines, LTE analysis
overestimates the abundance by $0.15-0.35$\,dex, depending on the line
strength and stellar parameters. As seen in Fig.\,\ref{fig:NaT}, the
non-LTE corrected abundances agree reassuringly well within the
uncertainties.

The Mg abundance was determined from the Mg line at 571.1\,nm, for
which \citet{Shimanskaya00} report non-LTE corrections that are
smaller than 0.02\,dex for the relevant stellar parameters. We note
that adopting a calculated \citep[e.g.][]{Chang90}, rather than
experimentally determined f-value \citep{Lincke71} for this line would
systematically lower the Mg abundances by $\sim 0.1$\,dex. The doublet
lines of neutral Al at 669.6/669.8\,nm were used to infer the Al
abundance. We did not find any non-LTE calculations in the literature
for these lines in metal-poor giant stars, but since they are weak,
high-excitation lines, they should at least be more reliable abundance
indicators in LTE than the stronger UV resonance lines at
396.1/394.4\,nm \citep[e.g.][]{Asplund05}. Furthermore, since our
targets all have quite similar stellar parameters, the differential
non-LTE effects should be small.

\begin{figure*}
        \centering \includegraphics[width=13cm,angle=90,viewport=0cm
          2cm 18cm 23cm]{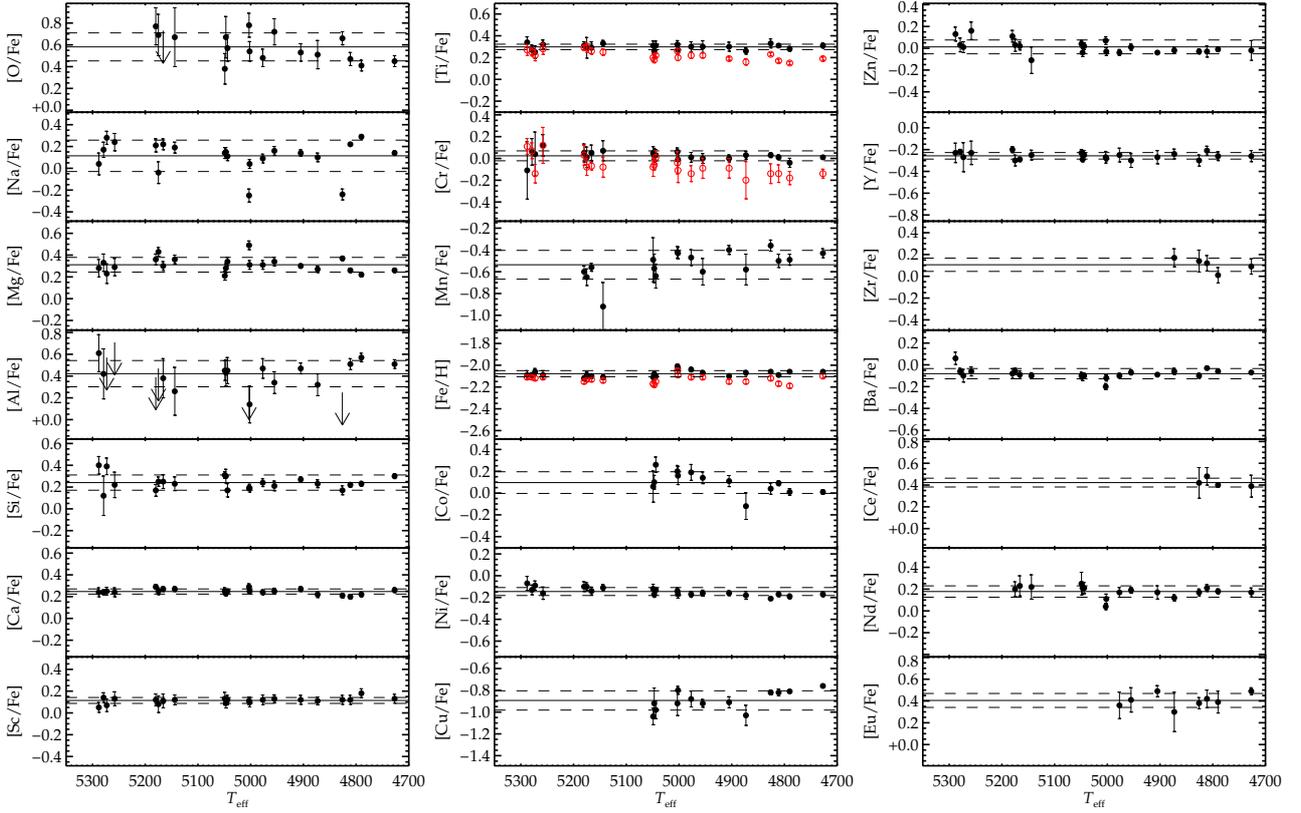}
          \caption{Elemental abundances of all targets. Each panel
            spans 1\,dex in abundance and marks the average abundance
            (solid line) and the $1\sigma$ dispersion limits (dashed
            lines). Two different symbols are used for [Ti/Fe],
            [Cr/Fe], and [Fe/H]: the red open circles and the black
            bullets in these panels represent abundances
            inferred from neutral and singly ionised lines,
            respectively. The hottest star in each panel marks the
            completeness limit, i.e. all sample stars cooler than this
            limit are shown.}
        \label{fig:AbvsT}          
\end{figure*}

\begin{table*}
\tiny
      \caption{Mean abundances and abundance dispersions compared to
        literature values for NGC\,6397. Listed are also the mean
        abundances of the RGB stars characterised as belonging to the
        first generation ($<A\rm(X)>_{I}$, 3 stars) and second
        generation ($<A\rm(X)>_{II}$, 18 stars), as well as the
        difference between the two mean quantities.}
      \label{tab:mabund}
      \centering
\begin{tabular}{lrrrrrrrrrrr}
\hline
               & $n_{\rm stars}$ & $n_{\rm lines}$  &  $<A\rm(X)>$ &$\sigma$ & C09$^{a,f}$&   K07$^{b,f}$ & G01$^c$ &C00$^d$ & $<A\rm(X)>_{I}$  & $<A\rm(X)>_{II}$ & $\Delta A\rm(X)_{\rm II-I}$  \\
\hline\hline                                                                    
$\rm[Fe/H] $     &  21  & 11--13&  $-2.08$  & $0.02$  & $-1.95$&    $-2.12$ & $-2.05$ &$-2.0$& $ -2.05 \pm0.02$ &$-2.08 \pm0.01$&$ -0.03\pm0.02$\\
$\rm[O/Fe]^e$    &  17  & 1--3  &  $ 0.58$  & $0.13$  & $0.35 $&    $...  $ & $0.37 $ &$0.15$& $  0.71 \pm0.04$ &$ 0.56 \pm0.03$&$ -0.15\pm0.05$\\
$\rm[Na/Fe]$     &  21  & 1--2  &  $ 0.11$  & $0.14$  & $0.11 $&    $...  $ & $0.31 $ &$0.19$& $ -0.18 \pm0.07$ &$ 0.16 \pm0.02$&$  0.34\pm0.07$\\
$\rm[Mg/Fe]$     &  21  & 1     &  $ 0.31$  & $0.07$  & $0.25 $&    $0.37 $ & $0.20 $ &$... $& $  0.43 \pm0.03$ &$ 0.29 \pm0.01$&$ -0.14\pm0.04$\\
$\rm[Al/Fe]^e$   &  21  & 1--2  &  $ 0.42$  & $0.12$  & $...  $&    $...  $ & $0.35 $ &$... $& $  <0.34       $ &$ 0.44 \pm0.03$&$  0.10\pm0.07$\\
$\rm[Si/Fe]$     &  21  & 2--4  &  $ 0.24$  & $0.07$  & $0.28 $&    $...  $ & $...  $ &$0.27$& $  0.20 \pm0.02$ &$ 0.25 \pm0.02$&$  0.04\pm0.03$\\
$\rm[Ca/Fe]$     &  21  & 16    &  $ 0.25$  & $0.02$  & $...  $&    $0.37 $ & $...  $ &$0.20$& $  0.25 \pm0.02$ &$ 0.25 \pm0.01$&$ -0.00\pm0.02$\\
$\rm[Sc/Fe]$     &  21  & 5--6  &  $ 0.11$  & $0.03$  & $...  $&    $...  $ & $...  $ &$... $& $  0.10 \pm0.01$ &$ 0.12 \pm0.01$&$  0.01\pm0.01$\\
$\rm[Ti/Fe]$     &  21  & 3--5  &  $ 0.30$  & $0.03$  & $...  $&    $0.22 $ & $...  $ &$0.36$& $  0.31 \pm0.01$ &$ 0.30 \pm0.01$&$ -0.02\pm0.01$\\
$\rm[Cr/Fe]$     &  21  & 1--2  &  $ 0.02$  & $0.05$  & $...  $&    $...  $ & $...  $ &$... $& $  0.03 \pm0.01$ &$ 0.02 \pm0.01$&$ -0.01\pm0.02$\\
$\rm[Mn/Fe]$     &  17  & 1     &  $-0.54$  & $0.13$  & $...  $&    $...  $ & $...  $ &$... $& $ -0.48 \pm0.09$ &$-0.55 \pm0.03$&$ -0.07\pm0.09$\\
$\rm[Co/Fe]$     &  13  & 1     &  $ 0.10$  & $0.10$  & $...  $&    $...  $ & $...  $ &$... $& $  0.12 \pm0.08$ &$ 0.09 \pm0.03$&$ -0.03\pm0.09$\\
$\rm[Ni/Fe]$     &  21  & 8--13 &  $-0.14$  & $0.04$  & $...  $&    $...  $ & $...  $ &$... $& $ -0.15 \pm0.03$ &$-0.14 \pm0.01$&$  0.01\pm0.03$\\
$\rm[Cu/Fe]$     &  13  & 1     &  $-0.89$  & $0.09$  & $...  $&    $...  $ & $...  $ &$... $& $ -0.87 \pm0.05$ &$-0.90 \pm0.03$&$ -0.03\pm0.06$\\
$\rm[Zn/Fe]$     &  21  & 1     &  $ 0.01$  & $0.06$  & $...  $&    $...  $ & $...  $ &$... $& $  0.02 \pm0.03$ &$ 0.01 \pm0.02$&$ -0.01\pm0.03$\\
$\rm[Y /Fe]$     &  21  & 2     &  $-0.26$  & $0.03$  & $...  $&    $...  $ & $...  $ &$-0.16$&$ -0.29 \pm0.01$ &$-0.25 \pm0.01$&$  0.04\pm0.01$\\
$\rm[Zr/Fe]$     &  5   & 1     &  $ 0.11$  & $0.06$  & $...  $&    $...  $ & $...  $ &$... $& $  0.14 \pm0.10$ &$ 0.10 \pm0.03$&$ -0.04\pm0.11$\\
$\rm[Ba/Fe]$     &  21  & 1--2  &  $-0.08$  & $0.05$  & $...  $&    $-0.18$ & $...  $ &$-0.16$&$ -0.12 \pm0.04$ &$-0.07 \pm0.01$&$  0.05\pm0.04$\\
$\rm[Ce/Fe]$     &  4   & 1     &  $ 0.42$  & $0.04$  & $...  $&    $...  $ & $...  $ &$... $& $  0.42 \pm0.14$ &$ 0.42 \pm0.03$&$  0.00\pm0.14$\\
$\rm[Nd/Fe]$     &  16  & 3--5  &  $ 0.18$  & $0.05$  & $...  $&    $...  $ & $...  $ &$... $& $  0.14 \pm0.05$ &$ 0.19 \pm0.01$&$  0.05\pm0.05$\\
$\rm[Eu/Fe]$     &  8   & 1     &  $ 0.40$  & $0.06$  & $...  $&    $...  $ & $...  $ &$... $& $  0.38 \pm0.05$ &$ 0.41 \pm0.03$&$  0.03\pm0.06$\\
\hline                                                                                                                                                
\multicolumn{6}{l}{$^a$ \citet{Carretta09b}, mean of 13 targets. }  & \multicolumn{5}{l}{$^d$ \citet{Castilho00},}     \\ 
\multicolumn{6}{l}{[Fe/H] based on FeI lines.}                       & \multicolumn{5}{l}{mean of ten brightest targets.}            \\
\multicolumn{6}{l}{$^b$ \citet{Korn07}, mean of six brightest targets.}   &  \multicolumn{5}{l}{$^e$ Upper limits treated as detections.}  \\
\multicolumn{6}{l}{$^c$ \citet{Gratton01}, mean of three brightest targets.} &\multicolumn{5}{l}{$^f$ \citet{Asplund09} solar abundances adopted.}
 \\

\end{tabular}
\end{table*}

\section{Abundance analysis}
\label{sec:analysis}
Fig.\,\ref{fig:AbvsT} shows [Fe/H] and abundance ratios relative to
Fe\footnote{$[X/Y]=\log{\left(\frac{N(X)}{N(Y)}\right)}_{*}-\log{\left(\frac{N(X)}{N(Y)}\right)}_{\sun}$
} for all stars (see individual star abundances in Appendix A). 
Table \ref{tab:mabund} lists the mean abundance and
dispersion in all elements, compared to values found in the
literature. Only recent studies that analysed stars on the upper RGB,
in approximately the same magnitude range as our targets, are included
in the comparison.   

\begin{figure}
        \centering
                \includegraphics[width=4.3cm,angle=90,viewport=1cm 0cm 10cm 25cm]{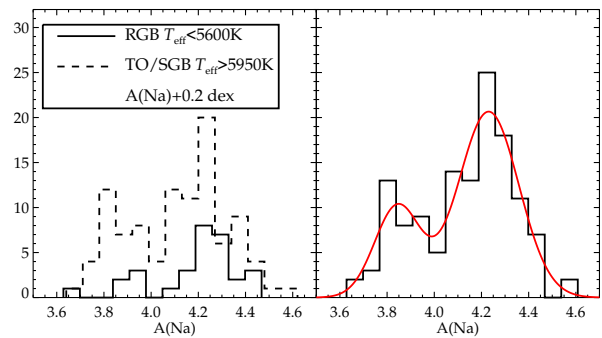}
          \caption{The left hand panel shows two histograms of the
            abundances displayed in Fig.\,\ref{fig:NaT}, separating RGB stars
            (solid line) from TO and SGB stars (dashed line). A shift
            of $+0.2$\,dex has been added to the TO/SGB group to
            correct for the presumed effects of atomic diffusion (see
            text). The right hand panel shows a histogram of the whole
            sample and a double-Gaussian fit to the distribution. } 
        \label{fig:NaHist}                                                                                       
\end{figure}

\subsection{O, Na, Mg, and Al}
\label{sec:light}
In agreement with previous findings for the cluster \citep[e.g.][and
  Paper I]{Carretta05,Carretta09b}, we find a large spread in Na
abundance. Fig.\,\ref{fig:NaT} shows how Na abundances vary with
effective temperature for post-TO stars in the cluster, including the
RGB star abundances determined in this study, as well as our improved
abundance determinations for the sample of TO, SGB, and RGB stars
analysed in Paper I.  It is apparent that the mean Na abundance is
higher by approximately 0.2\,dex in the RGB stars ($T_{\rm
  eff}<5600$\,K), compared to the TO and SGB group ($T_{\rm
  eff}>5950$\,K). For these low-mass stars, classical models do not
predict any dredge-up of fresh Na produced in situ by proton-capture
on pristine $^{22}$Ne \citep[see e.g.][]{Charbonnel10}, so the reason
for the apparent Na enhancement must lie elsewhere. A similar
difference was identified by \citet{Dorazi10}, between dwarfs and
giants in 47\,Tuc, which they propose to be an artificial result,
stemming from the assumption of Gaussian line profiles when
determining the equivalent widths of the 818.3/819.4\,nm lines. This
is not the case for our study, since we use a full spectrum
synthesis. Considering in particular the good agreement between the
818.3/819.4\,nm lines and the 568.2/568.8\,nm lines for the five RGB
stars for which both are observed (the mean difference between the
doublet is $0.03\pm0.04$, we propose that the 0.2\,dex difference is
  real. This is indeed the amount expected from stellar structure
  models including atomic diffusion, moderated by a certain degree of
  turbulence below the outer convection zone, as illustrated in
  Fig.\,\ref{fig:NaT} \citep[see][and references
    therein]{Richard05a}. \citet{Korn07} and \citet{Lind08} found
  differences of similar size between dwarf and giant star abundances
  of Fe and Mg in NGC\,6397, that are well reproduced by the T6.0
  model of \citeauthor{Richard05a}. We therefore propose that the TO
  and SGB stars have lower photospheric Na levels because of
  gravitational settling from the envelope, whereas the initial
  abundance has been reinstated in the photospheres of RGB stars, by
  the dredge-up of segregated pristine material when the convection
  zone deepened. We note that even if atomic diffusion accounts
  accurately for the abundance differences between dwarfs and giants
  in NGC\,6397, the situation may be different for 47\,Tuc. This
  cluster has a higher metallicity ([Fe/H]=-0.76, \citealt{Koch08})
  and hosts TO stars with cooler effective temperatures and larger
  convective envelopes, which implies that settling is less efficient
  in this cluster.

Accounting for the presumed effect of atomic diffusion and adjusting
the Na abundances of TO and SGB stars by $+0.2\,$dex, the abundance
histograms shown in Fig.\,\ref{fig:NaHist} reveal an interesting
pattern. The Na abundance distribution of both groups clearly appears
to be bimodal, having two peaks separated by $\approx0.4$\,dex and the
one at higher abundance being larger than the other. The right hand
panel of Fig.\,\ref{fig:NaHist} shows the histogram of the merged
sample, where the two peaks have become more distinguished. A
two-population fit, assuming Gaussian distributions, is also indicated
by a red solid line. The best fit places the first population around
$A\rm(Na)=3.84\pm0.10$ (mean and standard deviation) or
correspondingly $\rm[Na/Fe]\approx -0.30$, and the second around
$A\rm(Na)=4.23\pm0.13$ or $\rm[Na/Fe]\approx 0.09$. The number ratio
is approximately 1:3, with 25\% of the stars in the first group and
75\% in the second group. We interpret this abundance distribution as
a result of intra-cluster pollution, raising the Na abundances of a
second generation of stars relative to the first. This clear
distinction between the present populations was impossible before for
NGC\,6397, probably because of small number statistics. A one-sided
Anderson-Darling test of the cumulative distribution function returns
a probability of 11\% that the Na abundances can be described by a
single normal distribution with the mean and standard deviation of the
full sample ($<A\rm(Na)>=4.17$ and $\sigma=0.21$).

\begin{figure}
        \centering \includegraphics[width=7cm,angle=90]{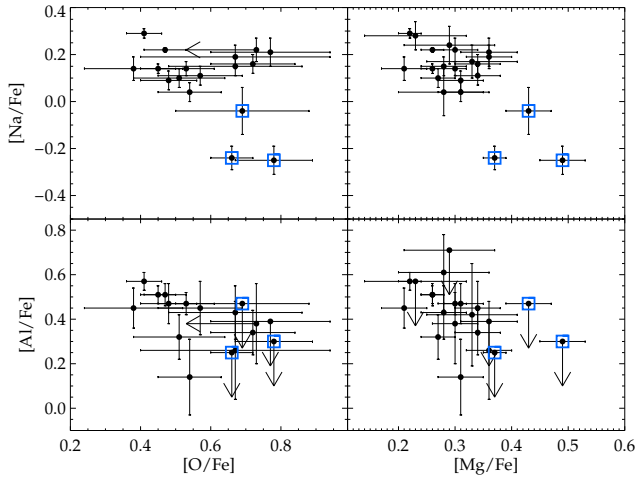}
          \caption{Na, Mg, and Al abundances for 21 stars and O
            abundances for 17 stars. The three Na-poor stars that we
            assign to the first generation are marked with by blue
            squares.}
        \label{fig:antis}                                                                                    
\end{figure}

Two of the RGB stars analysed in this study have Na abundances that
clearly place them in the first generation ($A\rm(Na)<4.0$, \#5644 and
\#17691). One additional star may be regarded as a limiting case
($A\rm(Na)=4.10\pm0.09$, \#14592), but after considering its
comparably high Mg and O levels we also assign this star to the first
generation.  The remaining 18 RGB stars are assigned to the second
generation. We ascribe to low number statistics the finding that only
three out of 21 RGB stars belong to the first generation (since the
number ratio found for the full sample is approximately $1:3$, we
would have expected five stars). As can be seen from
Fig.\,\ref{fig:antis}, displaying the O, Na, Mg, and Al abundances,
the populations are most clearly distinguished in a Na--Mg diagram,
probably because of the more precise determination of these abundance
ratios compared to O and Al. The Al lines are not detectable in the
first generation stars and we can only infer upper limits to their Al
abundance. Nevertheless, the abundances of O, Al, and Mg support the
view that the Na-rich stars have been polluted by gas enriched in
H-burning products. Table \ref{tab:mabund} lists the mean abundance
ratios relative to iron of the stars that we characterise as belonging
to the first and second generation. The difference in mean abundance
ratios is also listed.

\citet{Carretta09a,Carretta09b} measured the Na abundances of a large
number of RGB stars in NGC\,6397 and reported O abundances or upper
limits for a subset of 13 stars. Table \ref{tab:mabund} lists the mean
[O/Fe], [Na/Fe], and [Mg/Fe] ratios for this subset, using the same
solar abundances as in this study \citep{Asplund09}. Accounting for
the $\sim0.14$\,dex difference in mean [Fe/H] between the studies, the
mean O, Na, and Mg abundances agree rather well
($\pm0.1$\,dex). However, the full set of Na abundances published by
Carretta et al. (including the lower resolution measurements with
VLT/GIRAFFE) shows a systematic offset towards higher values by
typically 0.2\,dex. The reason for this offset can be traced to
differences in the adopted oscillator strengths ($\sim 0.05$\,dex), as
well as differences in the non-LTE corrections applied ($\sim
0.15$\,dex).

According to the distinction recognised by Carretta et al.\ the
primordial (first) generation consists of stars with
$\rm[Na/Fe]\le[Na/Fe]_{min}+0.3\,dex$, where $\rm[Na/Fe]_{min}$ is the
minimum value observed in the cluster and 0.3\,dex corresponds to
$\sim 4 \sigma$[Na/Fe] and $\sigma$[Na/Fe] is the star-to-star error
in [Na/Fe]. Adopting this definition for our RGB sample, the first
generation would consist of the same three stars,
($\rm[Na/Fe]_{min}=-0.25$), but barely overlap to also include two of
our second-generation targets. All the other sample stars would be
considered as second-generation stars.  Furthermore, Carretta et
al.\ define an extreme component among the second generation stars, if
present, as stars having $\rm[O/Na]<-0.9$ (or $\rm[O/Na]<-1.03$,
adopting the solar abundances of \citealt{Asplund09}). For NGC\,6397,
none of the 13 targets analysed by Carretta et al.\ fall in the
extreme component according to this definition and the same is true
for our sample. In numbers, Carretta et al.\ find that 25$\pm$13$\%$
and 75$\pm$22$\%$ of their full sample of stars in NGC\,6397 belong to
the primordial and intermediate components, respectively, which agrees
perfectly with our results (25\% and 75\%).
 
Finally, we have also inspected the Na abundance histogram of the full
Carretta et al.\ (2009) sample, to see whether the same double-peaked
distribution as in Fig.\,\ref{fig:NaHist} can be distinguished. This
is not the case however, which we believe is likely due to the larger
measurement errors (typically $\sim0.1$\,dex, according to their
estimate, compared to $\sim0.05$\,dex in our study) stemming from
their lower spectral quality. The possibility that the bimodality seen
in our abundance data is artificial and not representative of the true
distribution obviously cannot be completely ruled out. This is
unlikely though, especially considering that the two samples (TO+SGB
and RGB) display a very similar pattern. Furthermore, our Na abundance
histogram for NGC6397 bears a striking resemblance to the one found by
\citet[][see their Fig.\,6]{Marino08} for the higher metallicity
cluster M4. On the basis of this histogram, the authors find that
$\sim36\%$ of the stars are centred around $\rm[Na/Fe]=0.07\pm0.01$,
whereas the remaining majority of stars are of second generation,
centred around $\rm[Na/Fe]=0.38\pm0.01$.

\begin{figure}
        \centering \includegraphics[width=7cm,angle=90]{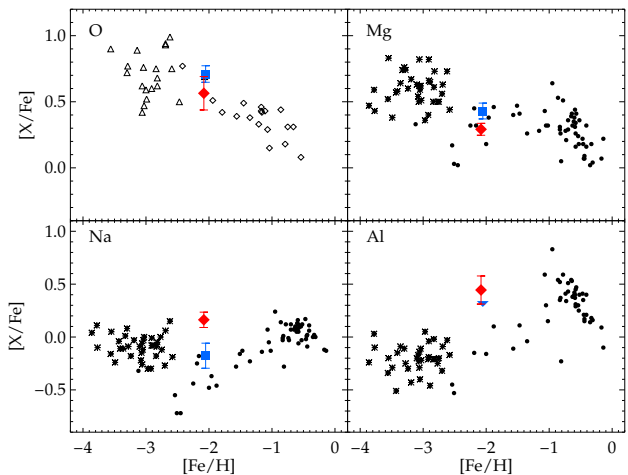}
          \caption{Comparison between abundances of RGB stars in
            NGC\,6397 and halo field stars. The blue filled squares
            represent the mean abundances of the NGC\,6397 stars
            assigned to the first generation and the red diamonds the
            mean of the second generation. The error bars mark the
            $1\sigma$ dispersion. Upper limits are treated as
            detections for the second generation averages for Al (two
            stars) and O (one star) when computing the mean and
            standard deviation. The literature data for halo field
            stars are assembled as follows; bullets represent
            \citet{Gehren06}, star symbols represent
            \citet{Andrievsky07,Andrievsky08,Andrievsky10}, open
            triangles \citet{Cayrel04}, and open diamonds
            \citet{Nissen02}.}
        \label{fig:field}  
\end{figure}

Fig.\,\ref{fig:field} compares the abundance ratios of our cluster
targets to halo field data. As expected, we see that the [Na/Fe] ratio
of the first generation is in good agreement with the field at this
metallicity, whereas the Na abundances of the second generation are
clearly enhanced. The [Al/Fe] ratios of the second generation also
appear enhanced relative to the field, but only upper limits can be
inferred for the first generation, limiting our comparison to field
stars. The [Mg/Fe] and [O/Fe] ratios have smaller total ranges in the
cluster than [Na/Fe] and [Al/Fe], which reflects that the absolute
abundances of Mg and O are higher than those of Al and Na by almost
three orders of magnitude. The measurements of O and Mg for 
both first and second generation stars may be compatible with those of
the field.

\subsection{Nitrogen}
\label{sec:nitrogen}
\begin{figure}
        \centering \includegraphics[width=14cm,angle=90,viewport=0cm
          3cm 20cm 25cm]{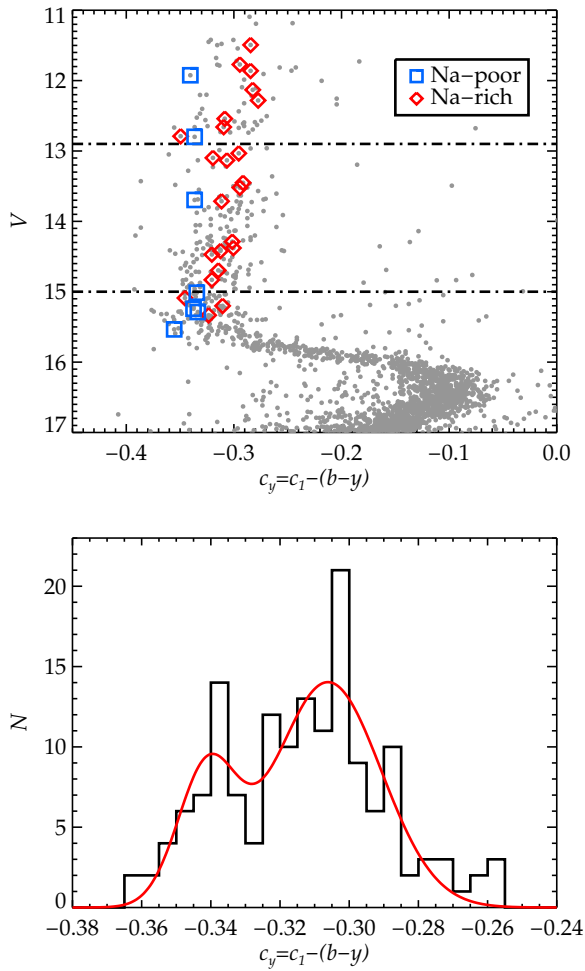}
          \caption{The top panel shows the $V-c_y$ diagram of
            NGC\,6397. Na-poor and Na-rich stars are indicated by blue
            squares and red diamonds respectively (abundance data from
            this study and Paper I). The bottom panel shows a
            histogram of the $c_y$ index for red giants in the
            magnitude range $12.9<V<15$. For details, we refer to the
            main text.}
        \label{fig:CyHist}                                                                                       
\end{figure}

As mentioned in the introduction, NGC\,6397 photometric data appear to
be very homogeneous. However, colour indices designed specifically to
trace elemental abundances may provide further support for the
presence of multiple populations. As realised by \citet{Grundahl02b},
the strength of the Str\"{o}mgren $c_1$-index can be used as a tracer
of the N abundance, since the $u$-filter involved in constructing this
index covers the 336\,nm NH features. \citet{Yong08} empirically
defined a more suitable index for the purpose, $c_y=c_1-(b-y)$, which
removes much of the temperature sensitivity of $c_1$. The top panel of
Fig.\,\ref{fig:CyHist} shows a $V-c_y$ diagram, in which we have
marked the first and second generation targets in this study with
different symbols. We have also added the RGB targets analysed in
Paper I, assigning stars with $A\rm(Na)<4.0$ to the first generation
and $A\rm(Na)>4.0$ to the second. The clear separation between the
generations visible in Fig.\,\ref{fig:CyHist} indicates that their N
abundances are also affected. This correlation between $c_y$-index and
light element abundances was also illustrated by \citet{Milone10} for
the more metal-rich globular cluster NGC\,6752. In both clusters, the
first generation of stars seem to populate a tight, blue RGB sequence
in this colour index, and the second generation a more dispersed, red
sequence. We also note the close analogy with the divided Na-rich and
Na-poor RGB sequences in the $U-(U-B)$ diagram of M4, as shown by
\citet{Marino08}.

In the bottom panel of Fig.\,\ref{fig:CyHist}, we have extracted a box
centred on the middle RGB and produced a histogram of the $c_y$
index. The magnitude cuts are applied to reduce any remaining bias due
to temperature effects. We again see a two-peaked histogram, very
similar to the one obtained for the Na abundance. The best fit returns
the peak positions at $c_y=-0.341$ and $c_y=-0.306$, with a size
relationship of 25\% and 75\% for the left and right peak, in
excellent agreement with previously inferred number fractions for the
first and second stellar generation (see Sect.\,\ref{sec:light}).
However, the probability that the colours can be described by a single
Gaussian distribution is 59\%, again as given by the Anderson-Darling
statistics.

We can estimate the corresponding difference in [N/Fe] abundance using
the theoretical calculations (based on synthetic colours) by
\citet{Onehag09}, giving an approximate sensitivity of $\delta
c_y/\delta\rm[N/Fe]=0.04\rm $ for stars in the same $T_{\rm eff}$ and
$\log{g}$-ranges in NGC\,6752. A somewhat stronger sensitivity, 0.06,
was found by \cite{Yong08} who empirically correlated spectroscopic N
abundances with $c_y$-colour for the same stars. This cluster has a
slightly higher metallicity, but still helps us to obtain an order of
magnitude estimate. The typical $\Delta\rm[N/Fe]$ between stars of the
first and second generation thus lies in the range 1--1.5\,dex. This
is consistent with spectroscopic N measurements by \citet{Carretta05},
who found that six out of nine SGB stars are strongly enhanced in
[N/Fe] compared to the remaining sample, by as much as
1.5\,dex. \citet{Pasquini08} find all three stars in their sample to
be highly N-enriched relative to the field. Spectroscopic N
measurements for a large sample of stars in NGC\,6397 is obviously
desirable to establish the absolute range of abundances and to verify
the apparent bimodality.
 
\subsection{$\alpha$ and iron-peak elements}
\label{sec:medium}
Globular clusters are generally very homogeneous in $\alpha$ and
iron-peak elements, with the exceptions of peculiar clusters such as
$\omega$ Cen. In NGC\,6397, the star-to-star scatter in [Fe/H] in our
RGB sample is extremely small, only $0.03\,$dex, when inferring the
iron abundance from singly ionised lines. The Fe\,I line abundances
exhibit a slight decreasing trend with decreasing effective
temperature, which is not seen in Fe\,II line abundances. The same
pattern is seen for [Ti/Fe] and [Cr/Fe] ratios, and may indicate that
the temperature scale contains a small systematic bias ($\sim50$\,K
over a total range of 500\,K) or that non-LTE and/or 3D effects have
minor differential effects. In Table \ref{tab:mabund}, the reported
Fe, Ti, and Cr abundances are all based entirely on singly ionised
lines. The abundances of Mn, Co, Cu, and Zn are based on one rather
weak neutral line, which likely explains the larger star-to-star
scatter in these abundances.

In general, the $\alpha$ and iron-peak elements display the expected
pattern, in agreement with the halo field at this metallicity. The
abundances of Ca, Si, Ti, and Sc are overabundant by $0.1-0.3$\,dex,
whereas the iron-peak elements Cr, Co, Ni, and Zn are approximately
solar. The odd-z elements Mn and Cu are strongly under-abundant
relative to the Sun, and the low value we inferred for Cu
($\rm<[Cu/Fe]>=-0.89$) is slightly below the values found for field
dwarfs at this metallicity by \citet{Primas08}. The mean Mn abundance
($\rm<[Mn/Fe]>=-0.54$) is very similar to the LTE abundances inferred
by \citet{Bergemann08}, but as shown in that study, non-LTE
corrections for Mn are positive and rather substantial.

As can also be seen in Table \ref{tab:mabund}, there is no significant
difference (i.e. $>2\sigma$) between the mean abundances of
intermediate mass (heavier than Al) and iron-peak elements for the
stars that we characterise as belonging to the first and second
generation.

\subsection{Neutron capture elements}
\label{sec:heavy}
Metal-poor halo field stars are known to display a large scatter in
some of the neutron capture elements, which cannot be explained by
observational uncertainties \citep[see
  e.g.][]{Barklem05,Burris00}. Stars within a globular cluster are
generally far more homogeneous in their heavy element abundances, as
shown by for example \citet{James04} for NGC\,6397, NGC\,6752, and
47\,Tuc and by \citet{Yong08b} for M4 and M5. This is confirmed by our
study, in which small abundance dispersions are found for Y, Zr, Ba,
Ce, Nd, and Eu. The mean abundance ratios we derive for
$\rm[Y/Fe]=-0.26$ and $\rm[Ba/Fe]=-0.08$ are in reasonable agreement
with previous estimates for the cluster (in addition to the values
listed in Table \ref{tab:mabund}, James et al.\ determined
$\rm[Y/Fe]=-0.2$ and $\rm[Ba/Fe]=-0.17$ for stars at the base of the
RGB). The other elements have not been previously analysed in
NGC\,6397. By comparing to the values determined for the higher
metallicity cluster NGC\,6752 by \citet{Yong05}, we can observe good
agreement between Zr, Ba, Ce, Nd, and Eu abundances. However, for Zr
this is partly coincidental; accounting for differences in the adopted
oscillator strength for the Zr\,II line at 511.2\,nm, our abundance
ratio would be 0.3\,dex lower. A similar difference is seen when
comparing the estimates for [Y/Fe]. It thus seems that the two
clusters are very similar in heavy neutron-capture elements having
$\rm Z>56$, whereas the lighter $\rm Z=39-40$ elements are 0.3\,dex
less abundant in NGC\,6397.

According to \citet{Burris00}, Y, Zr, Ba, and Ce are all mainly
s-process elements (72-85\% s-process contribution), whereas Nd is
approximately half r-process, half s-process. Eu is an almost pure
(97\%) r-process element. For the first time, we determine the [Ba/Eu]
ratio in this cluster, and the value we infer ($-0.49$) is quite close
to the scaled solar r-process of $-0.81$ (the scaled s-process value
is $1.45$). The same is true for the [Zr/Eu], [Ce/Eu], and [Nd/Eu]
ratios, whereas [Y/Fe] is even slightly below the value expected for a
pure r-process contribution. This implies that chemical enrichment by
core-collapse supernovae may be sufficient to explain the observed
heavy element abundances, and that the contribution from AGB stars has
been limited.

The two stellar generations are very similar in terms of the measured
neutron-capture elements (see Table \ref{tab:mabund}). Only the [Y/Fe]
ratio is higher by $\sim3\sigma$ in the second generation. The
difference (0.04\,dex) is, however, too small to be convincing. A
comparison with halo field stars does not provide much information,
since the stars in the field are highly scattered. The abundances we
find for NGC\,6397 are well within the ranges found e.g. by
\citet{Francois07} for a sample of metal-poor field giants.

\subsection{Helium}
\label{sec:helium}
If the second generation stars in globular clusters have indeed been
enriched in nucleosynthesis products connected to hydrogen-burning,
they should also have a higher He content than their first generation
counterparts, although we may only speculate about the exact amount.

\citet{diCriscienzo10} used the accurate, proper-motion cleaned HST
photometry of \citet{Richer08} to measure the width of the main
sequence in NGC\,6397 and constrain the possible helium spread in the
cluster. Under the assumption that 70\% of the stars belong to the
second generation and have a C+N+O enhancement as suggested by
spectroscopic observations ($\Delta\rm[(C+N+O)/Fe] \sim 0.25$
according to Carretta et al.\ 2005), they conclude that the cluster
must be very uniform in He abundance, implying that there is a
difference of only $\Delta Y=0.02$ between the first and second
generations. When allowing for higher [(C+N+O)/Fe] overabundances, a
broader spread in helium ($\Delta Y$ of up to 0.04) is compatible with
the tightness of the main sequence. Here we attempt to constrain this
difference using spectroscopic arguments instead.

The He abundance cannot be measured directly in late-type stars, but
nevertheless influences the emerging stellar spectrum. The effects of
helium-enrichment in a cool stellar atmosphere has been investigated
by \citet{Stromgren82}, who found that changes in the
helium-to-hydrogen ratio of F type dwarfs affect the mean molecular
weight of the gas and have an impact on the gas pressure. However,
helium contributes only negligible to the line and continuous opacity
in a cool atmosphere and does not provide free electrons. Therefore,
\citeauthor{Stromgren82} concluded and demonstrated that a
helium-enrichment can be mapped merely as a shift in surface gravity,
i.e.\ a helium-enriched atmosphere is similar to a helium-normal
atmosphere with a higher surface gravity, in terms of temperature
structure and electron pressure structure. Stars with identical $T$
and $P_{\rm e}$ structures and the same ratio of metals-to-hydrogen
also have identical spectra. The calculation of the necessary shift in
surface gravity is given by the relation (see equivalent eq.\ 12 in
\citealt{Stromgren82})

\begin{equation}
   \log{g''}=\log{g'}+\log\left(\frac{(1+4\times
     y')(1+y'')}{(1+4\times y'')(1+y')}\right)
\end{equation}

\noindent where $\log{g'}$ is the surface gravity of the model with
helium-to-hydrogen ratio $y'=N_{\rm He}/N_{\rm H}$ and analogously for
the double-primed variables. We tested whether the relation also holds
for our RGB targets by computing two metal-poor MARCS atmospheric
models \citep{Asplund97} with $T_{\rm eff}=4700\rm\,K$. One model has
extreme helium-enrichment, $\log{g}=1.65$, and $y=1.0$ (corresponding
to helium mass-fraction of $Y=0.8$). The second model is
helium-normal, but has a higher surface gravity, $\log{g}=1.95$, and
$y=0.085$ ($Y=0.25$). All metal abundances relative to hydrogen are
the same in both models ($\rm[Fe/H]=-2.0$), which means that the
helium-rich model has a lower mass fraction of metals. The temperature
structure and electron density structure of the two model atmospheres
are very similar, as expected, and the derived elemental abundances
based on the models are the same to within $\sim$0.005\,dex. We thus
consider it safe to use the scaling with surface gravity of normal
composition models to mimic helium-enhancement.

\begin{table}
      \caption{Stellar parameters for \#5644 and \#14565 derived assuming a normal-star hydrogen and helium composition.}
         \label{tab:params}
         \centering
         \begin{tabular}{lrrrrr}
                \hline\hline
                 ID        & $T_{\rm eff}$  & $T_{\rm eff}$ &  $T_{\rm eff}$   & $\log{g}$   & $\log{g}$    \\
                           & $b-y$          & H$\alpha$     &  Ex.\,eq.\,      & Photo.      & Ion.\,eq.\,     \\
                                 \hline 
                 5644      & 4826           & 4740          &  4597            & 1.74        & 1.50            \\
                 14565     & 4811           & 4730          &  4598            & 1.71        & 1.52            \\
                 $\Delta$  &   15           &   10          &    -1            & 0.03        &-0.02            \\
                                 \hline
         \end{tabular}
\end{table}

We selected one star from each generation to test the hypothesis of a
difference in He-content. The stars have essentially identical stellar
parameters, which allows a differential comparison. Table
\ref{tab:params} lists the derived stellar parameters for these two
targets. The absolute $T_{\rm eff}$-values inferred from fitting the
wings of the H$\alpha$ line and from the excitation equilibrium of
neutral iron are both somewhat lower than the values derived from the
photometric data, but the difference between the two selected stars is
small for all methods. Since the true difference in He abundance
between the stars, if any, is unknown, the effective temperatures have
been computed by assuming a normal helium-to-hydrogen ratio in both
objects. None of the methods should be severely biased however by this
assumption. The Str\"{o}mgren colours have only a weak sensitivity to
surface gravity ($\sim\pm15$\,K per $\pm0.1$\,dex in $\log{g}$), which
is also true for neutral iron lines (the differential impact on the
iron abundance inferred from low- and high-excitation lines is
$0.001$\,dex per $0.1$\,dex in $\log{g}$, which translates to
$<0.5$\,K in effective temperature.). The influence of a higher
helium-to-hydrogen ratio on the H$\alpha$ profile has been
investigated and found to be negligible as well, if the number density
of metals remains constant. Based on the information in Table
\ref{tab:params}, we conclude that the difference in effective
temperature between the stars is likely to be small. We assume that
they have the same effective temperature and adopt 10\,K as a
representative error in the relative difference.

The photometric surface gravities were determined in Paper I, from the
common relation between mass, surface gravity, luminosity, and
temperature. Assuming $M=0.8\rm\,M_{\sun}$ and bolometric correction
$BC=-0.390$, the resulting values are different by 0.03\,dex
only. However, as the evolution of a helium-enhanced star is certainly
different from that of a helium-normal one, the assumption of equal
masses for the two stars may be wrong. If \#14565 is helium-enriched
compared to \#5644, it should, from stellar-evolution arguments, have
a lower mass and consequently a lower surface gravity. We also
calculated surface gravities based on the ionisation equilibrium of
Fe\,I and Fe\,II. With a normal-star composition for both objects, the
surface gravity values are different by 0.02\,dex. If a higher helium
mass fraction is assumed for 14565, the surface gravity value that
establishes the equilibrium will shift to lower values in accordance
with Eq.\, 1. However, the iron abundance found relative to hydrogen
will be the same in both cases. We thus conclude that if the stars
have the same helium mass fraction, they also have almost the same
surface gravity. If not, the difference in $\log{g}$ cannot be
determined independently from the difference in helium
content. However, if $\log{g}$ is calculated by ionisation
equilibrium, the effect of the helium enhancement will effectively
cancel and [Fe/H] will be found to be the same.

\begin{table}
      \caption{Iron mass fractions derived for \#5644 and \#14565,
        assuming different mass fractions of helium}
         \label{tab:helium}
         \centering
         \begin{tabular}{llllll}
                \hline\hline
                 ID    &   5644    &  14565      &  14565   &  14565        \\
                                \hline 
                 $y$   &   0.085    &  0.085       &  0.170    &   1.000    \\
                 $Y$   &   0.253    &  0.253       &  0.405    &   0.800    \\
		 $X$   &   0.747    &  0.747       &  0.595    &   0.200    \\	
                 $\log{g}$         & 1.50     & 1.52     &   1.45       & 1.21 \\
                 $A\rm(Fe)$        & 5.316    & 5.323    &   5.323      & 5.323  \\
                 $\rm Z\rm(Fe)*10^6$        & 8.662 & 8.803 &   7.007   & 2.353\\
                \hline
         \end{tabular}
\end{table}

We calculate the iron abundances, $A\rm(Fe)$, and iron mass fractions,
$\rm Z(Fe)$, of 5644 and 14565, using $T_{\rm eff}=4750\rm\,K$,
$\xi_{t}=1.7\rm\,km s^{-1}$, and the surface gravities that satisfy
the ionisation equilibrium of iron, which is different for different
values of $y$. The results are given in Table \ref{tab:helium}. When
assuming a normal He composition, the iron abundances of the stars
agree to within 0.007\,dex. A higher mass fraction of helium in 14565
will not affect the iron abundances derived relative to hydrogen, but
the mass fraction of iron will be lower. Assuming that the mass
fractions of iron are exactly the same for the stars, we can calculate
the corresponding difference in helium mass fraction. The errors in
$\Delta A\rm(Fe)$ and $\Delta\rm Z(Fe)$ are estimated from the
standard deviation in the mean difference for each individual line, to
cancel contributions from uncertainties in oscillator strengths. The
resulting error is 0.003\,dex for Fe\,I and 0.009\,dex for
Fe\,II. Adopting a relative uncertainty in effective temperature of
10\,K corresponds to an additional 0.016\,dex. The total error budget
is thus 0.028\,dex, or 6.7\% in iron mass fraction. Assuming that
\#14565 have the same $\rm Z(Fe)$ as \#5644, $\Delta Y$ between the
stars then becomes $0.01\pm0.06$. We caution that this is far from a
stringent constraint on the true difference in $Y$, especially
considering the possibility of a real (small) difference in iron mass
fraction. However, it is safe to say that our simple arithmetic
exercise clearly does not support extreme He-enhancement, in agreement
with the photometric tightness of the main sequence.

In the cluster formation scenario described by \cite{Decressin07b}, in
which slow winds from rotating massive stars pollute the star forming
gas, one expects an anti-correlation between helium and [O/Na] for
second generation stars born out of a mixture of stellar ejecta and
pristine material. As can be seen in Fig.\ 13 of Decressin et al.\,,
the small value implied for $\Delta Y$ is fully compatible with the
corresponding $\Delta\rm[O/Na]$ of 0.65~dex between \#5644 and
\#14565. This good agreement persists over the whole mass range
considered by \cite{Decressin07b}, i.e., for polluter stars with
masses between 20 and 120~M$_{\odot}$.

\section{Consequences for the evolutionary scenario of NGC\,6397}
\label{sec:conseq}
We now investigate whether the abundance patterns we derived for
NGC\,6397 can be accounted for in the framework of the so-called
``wind of fast rotating massive stars" (WFRMS) scenario \citep[see
  e.g.][]{Prantzos06,Decressin07b,Decressin07a}. In this scenario,
second generation stars are formed from matter ejected through slow
winds of individual massive stars (M$\geq$20~M$_{\odot}$), mixed
locally with pristine interstellar gas.

We note from the onset that the main alternative scenario, proposing
that ejecta from intermediate-mass AGB stars have shaped the
abundances patterns of light elements may be equally successful in
explaining the observed anti-correlation of O and Na, as well as an
accompanying mild He enrichment. In particular, \citet{Ventura09} show
how the O--Na data of \citet{Carretta05} can be reproduced if $50\%$
of AGB ejecta are mixed with $50\%$ pristine intracluster gas. The
main argument against such a scenario is the implied near constancy of
C+N+O, as discussed above. \citet{Decressin09} demonstrate how
metal-poor massive rotating AGBs are expected to produce very strong
C+N+O enhancement as He-burning products are also released in the
ejecta. In contrast, for the massive star scenario only H-burning
products are expected to enter the composition of second generation
stars, thus maintaining a stable sum of C+N+O.

\subsection{Method}
\label{sec:method}
We follow the same procedure as described in \citet{Decressin07b},
using the stellar models of \citet{Decressin07a} computed for the
metallicity of NGC\,6752 ([Fe/H]$\sim -1.56$), i.e. slightly higher
than that of NGC\,6397 ([Fe/H]$\sim -2.08$). A difference of this size
should have no impact on our conclusions \citep{Ekstrom08}. However,
the enrichment history of the two clusters certainly differ to some
extent, which must be considered in the analysis. In particular, the
full range in O abundance is twice as large in NGC\,6752 (0.8\,dex,
\citet{Yong03}) than in NGC\,6397 (0.4\,dex, see
Fig.\,\ref{fig:antis}). In addition, NGC\,6752 has a broadened main
sequence \citep{Milone10} and an extended horizontal branch,
indicative of a larger variation in the He content.

When establishing the recycling process of NGC\,6752,
\citet{Decressin07b} took into account all the material ejected by
massive stars rotating at critical velocity, both on the main sequence
and during the first part of the central He-burning
phase\footnote{Owing to strong mechanical mass loss the models predict
  that rapidly rotating stars actually evolve away from the critical
  limit during the central He-burning phase, before the He-burning
  products contaminate the slow wind component
  \citep[see][]{Decressin07a,Ekstrom08}.}. This allowed the build-up
of extreme abundance anomalies in both O and Na, and also lead to high
He-enrichment for the ``extreme component" of second generation
stars. As discussed in Sects.\,\ref{sec:light} and \ref{sec:helium},
such extreme anomalies appear to be lacking in NGC\,6397, and the
second generation stars are probably only mildly overabundant in
He. Consequently, we assume that only the slow winds ejected on the
main sequence by rapidly rotating massive stars are recycled into the
second generation and we neglect the material that is lost during the
luminous blue variable phase. As discussed in Sect.\,\ref{sec:mass},
this may have some impact on the total mass budget, although most of
the mass loss through the slow mechanical winds occurs while the
rapidly rotating stars are on the main sequence (in the case of a
60~M$_{\odot}$ rotating at critical velocity, 20~M$_{\odot}$ and
7~M$_{\odot}$ are lost through the mechanical wind on and slightly
after the main sequence, respectively, according to these models).  We
do not venture to discuss in detail the underlying reason for the
different observed abundance patterns in NGC\,6752 and NGC\,6397, but
one may speculate that different initial masses of these clusters have
played a role, especially on the amount of pristine gas available to
be mixed with the massive star ejecta to form second generation stars
(see Sect.\,\ref{sec:dilution}). We note also that cluster mass has
been identified as one of the main parameters governing the chemical
patterns of GCs, together with cluster metallicity, age \citep[see
  e.g.][]{Carretta09b,Carretta10b}, and environment of formation
\citep{Fraix-Burnet09}.

\subsection{Amount of dilution between massive star ejecta and pristine gas}
\label{sec:dilution}
\begin{figure}
  \centering \includegraphics[width=7cm,angle=90]{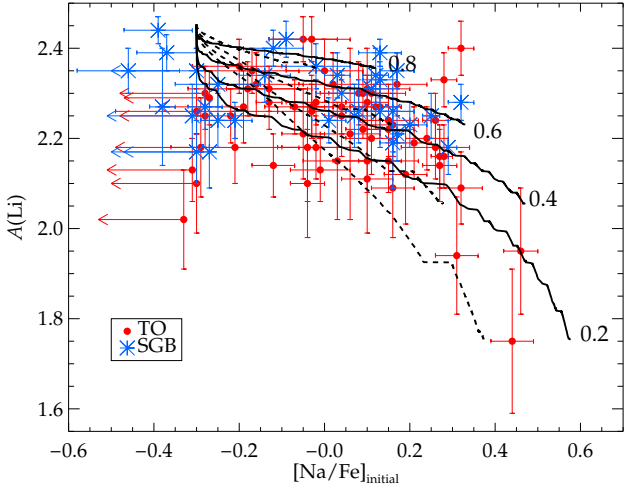}
  \caption{Li and Na abundances for TO and SGB stars in NGC\,6397
    (data from Paper I, with improved non-LTE derivation).  The
    initial [Na/Fe], shown on the abscissa, has been estimated by
    correcting $A\rm(Na)$ for atomic diffusion by $+0.2$\,dex and
    assuming $\rm[Fe/H]=-2.1$ for all stars. The red bullets represent
    turn-off stars and the blue stars subgiants.  Superimposed are the
    theoretical predictions for the chemical composition of the matter
    out of which second generation low-mass stars are expected to form
    in the WFRMS scenario. The tracks shown correspond to the cases of
    20~M$_\odot$ and 120~M$_\odot$ polluter stars (dashed and full
    lines, respectively) assuming different values (0.2, 0.4, 0.6,
    0.8) for the minimum dilution coefficient. See text for more
    details.}
  \label{fig:LiNamod}
\end{figure}

The star-to-star spread in light element abundances observed in
globular clusters today may imply that all the material ejected by the
first generation massive stars was not fully mixed before being
recycled into the second stellar generation. In addition, the
measurements of both O and Li pinpoints the formation of second
generation stars in the vicinity of individual massive stars from
their slow winds mixed with pristine interstellar matter.

As explained in \citet{Decressin07b} we may use the observed Li-Na
anti-correlation to estimate the amount of dilution between the
massive star ejecta that are totally void of Li and the pristine
intracluster gas left after the formation of first generation stars,
whose Li abundance is assumed to be equal to the primordial (Big Bang)
value. In addition, we assume that the slow wind of each individual
massive star experiences locally a variable dilution with time. In the
early main-sequence phase, the stellar ejecta encounter more pristine
gas and are hence more diluted than the matter ejected later. As the
stellar winds become more Na-rich later in the course of the evolution
of a massive polluter, the Li-Na anti-correlation builds up naturally:
a second generation of stars of a relatively high Li content and a
relatively low Na abundance are created first, while stars formed from
matter ejected later have less Li and more Na.

In Fig.\,\ref{fig:LiNamod}, we present theoretical predictions for the
Li and Na abundance variations in the resulting material out of which
second generation low-mass stars are assumed to form. The tracks shown
correspond to the cases of the ejecta of the 20~M$_\odot$ and
120~M$_\odot$ models by \citet{Decressin07a} for different values
(0.2, 0.4, 0.6, and 0.8) of the minimum dilution (along individual
tracks dilution decreases from 1.0 to this minimum value as the
polluter star evolves). Predictions for polluters with different
initial masses would fill the whole space between 20 and 120~M$_\odot$
tracks. The initial (pristine) [Na/Fe] value is assumed to be the
average that we find for first generation stars (i.e., $-0.3$; see
Sect.\,\ref{sec:light}). For Li, we consider the value from WMAP and
Big Bang nucleosynthesis ($A\rm(Li)=2.72$ according to
\citealt{Cyburt08}) and take into account Li depletion of 0.2\,dex at
the surface of both first and second generation low-mass stars during
the main sequence \citep[see][and Paper I]{Charbonnel05a}. In
Fig.\,\ref{fig:LiNamod}, the theoretical tracks are superimposed on
the Li and Na observational data by Paper I for NGC\,6397 TO and SGB
stars that have not yet undergone the first dredge-up. TO and SGB
stars (defined here as stars fainter and brighter than $V=16.2$,
respectively) are indicated by different symbols to also consider the
0.1\,dex difference in $A\rm(Li)$ found between these two groups,
which in Paper I was interpreted as a sign of a dredge-up of settled
Li. Various degrees of dilution can accurately explain the observed
spread, and the bulk of the sample can be reproduced assuming a
minimum dilution factor of 0.5 (i.e., that at least 50\% of the
material that formed the corresponding stars is pristine gas).  Stars
with low-Li and high-Na stars require a lower minimum dilution factor
of $\sim$ 0.2. Within the observational errors, the overall agreement
between the observational and theoretical trends and dispersions is
satisfactory.

\subsection{Expected anti-correlations}
\label{sec:anti}

\begin{figure}
   \centering \includegraphics[width=12cm,angle=90,viewport=0cm 13cm
     19cm 27cm]{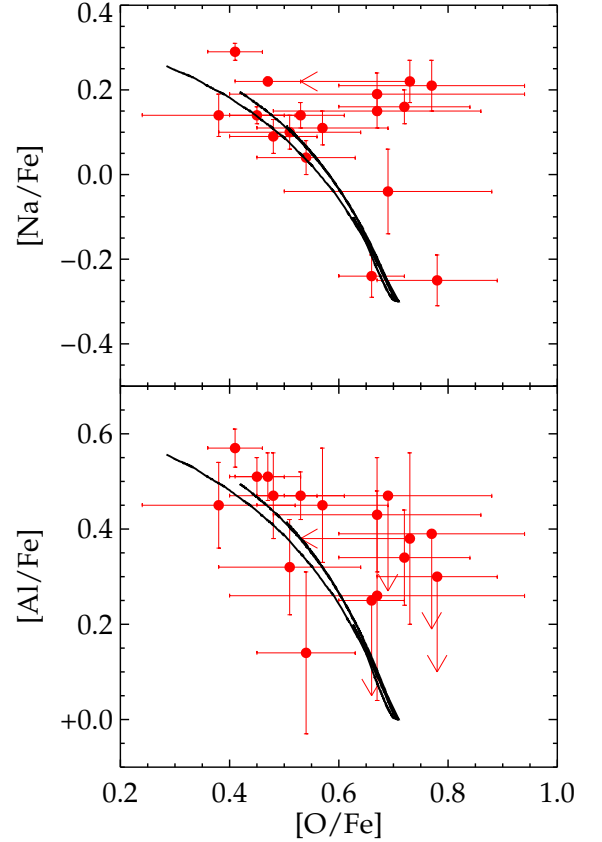}
  \caption{O-Na (top) and O-Al (bottom) anti-correlations are shown
    for the giant stars of NGC\,6397. Superimposed are the theoretical
    predictions for the chemical composition of the matter out of
    which second generation low-mass stars are expected to form. The
    tracks correspond to the cases of 20, 40, 60, and 120~M$_\odot$
    polluter stars for a minimum value of 0.2 for the dilution
    coefficient.}
  \label{fig:ONaAlmod}
\end{figure}

Armed with this calibration for the dilution between massive star
ejecta and pristine gas we can now make similar predictions for the
(anti-)correlations between O, Na, and Al. For the composition of the
pristine cluster gas, we use the average values that we derived for
first generation stars for O and Na (${\rm [O/Fe]}=0.7$, ${\rm
  [Na/Fe]}=-0.3$). Since we could only derive upper limits to the Al
abundance of the first generation, we assume a value that is
approximately compatible with the field at this metallicity,
${\rm[Al/Fe]}=0.0$ (see Fig.\,\ref{fig:field}).

The theoretical tracks for the chemical composition of material out of
which second generation stars can form are shown in
Fig.\,\ref{fig:ONaAlmod}. They correspond to the case of a minimum
dilution equal to 0.2 (for these elements, the tracks for other
minimal dilution values are superimposed on one another although to
different extents) for polluters of various initial masses. As can be
seen, the tracks do not vary much with polluter mass since the central
temperature in massive stars is always high enough to burn protons
through CNO-cycles and Ne-Na and Mg-Al chains. However, the efficiency
of the rotationally-induced mixing increases with stellar mass,
implying that more massive stars give rise to more extreme abundance
variations.  In particular, only the most massive stars can be held
responsible for the highest Al and Na enrichment observed in
NGC\,6397. Satisfactory agreement is again achieved between data and
predictions.

\citet{Decressin07a} described the sensitivity of the model
predictions to the nuclear reaction rates adopted in the
computations. The stellar models used here were computed with their
``set C" for the nuclear reaction rates \citep[see Table 2 of
][]{Decressin07a}. In this case, the total magnesium abundance
decreases by $\sim$ 0.1~dex only. If we were to assume an enhancement
of the $^{24}$Mg(p,$\gamma$) reaction as done in ``set D" of
\citet{Decressin07a}, we would however predict a total magnesium
abundance decrease by $\sim$0.15 dex, in slightly better agreement
with our Mg data in NGC\,6397.

\subsection{He content} 
\label{sec:content}

\begin{figure}
   \centering \includegraphics[width=7cm,angle=90]{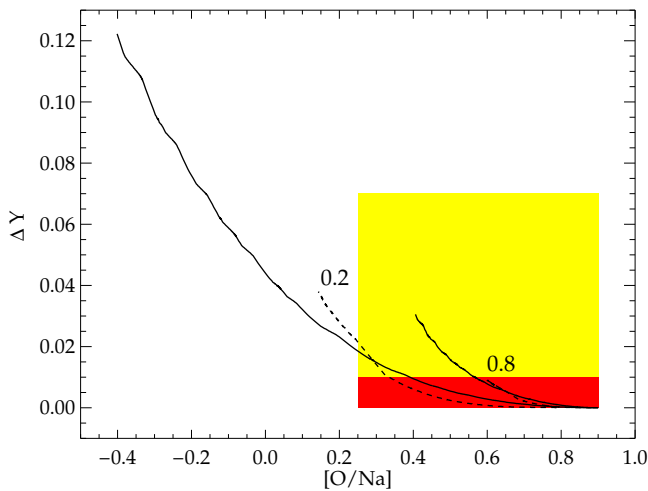}
  \caption{Expected anti-correlation between helium (in mass fraction)
    and [O/Na] for second generation stars (at birth) born out of
    material ejected through slow winds of massive stars and diluted
    with gas of pristine composition.  The tracks correspond to the
    cases of 20 and 120~M$_\odot$ polluter stars (dashed and full
    lines, respectively) assuming different values (0.2 and 0.8) for
    the minimum dilution coefficient. See text for more details. The
    shaded regions indicate the star-to-star abundance differences
    derived in Sect.\,\ref{sec:helium} for \#5644 and \#14565 and
    related uncertainties ($\Delta\rm[O/Na]=0.65$ and $\Delta\rm
    Y<0.07$). }
  \label{fig:ONavsHe}
\end{figure}

We can now make predictions about the helium spread expected within
the WFRMS scenario with the assumptions described above in the
specific case of NGC\,6397. Fig.\,\ref{fig:ONavsHe} shows the
predicted variation in He mass fraction as a function of the [O/Na]
ratio. We also show the values estimated in Sect.\,\ref{sec:helium}
for \#5644 and \#14565. We see that the theoretical helium variation
corresponding to a $\Delta$[O/Na] of 0.65 is smaller than 0.04, fully
compatible with the spectroscopic estimate, when very modest C+N+O
enrichment is considered as discussed above.

When assuming a ``standard" IMF (slope of 2.35) for the massive
polluter stars and taking into account the variable dilution process
described in Sect.\,\ref{sec:dilution}, we find that $\sim$ 70\% of
second generation stars should be born with an helium mass fraction
that is 0.02 higher than first generation stars. In addition, about
5-10\% of second generation stars are expected to be born within the
range $\rm Y=0.35-0.39$. The resulting dispersion in $Y$ is compatible
with the constraints obtained from the width of the main sequence
($\Delta Y<0.04$), although no blue, very He-rich main sequence has
been identified in the cluster. This prediction is, however, at odds
with the lack of a blue tail to the horizontal branch in NGC6397.

\subsection{Initial cluster mass}
\label{sec:mass}
As discussed in Sect.\,\ref{sec:light}, about 75\% of the long-lived
low-mass stars still present today in NGC\,6397 belong to the
so-called second generation, while about 25\% were born with the same
chemical composition as the massive polluters. \citet{Prantzos06} and
\citet{Decressin07b,Decressin07a} demonstrated that such high a number
ratio of second generation stars requires either a very flat IMF for
the first generation of massive stars, or that a very large number of
first-generation low-mass stars have escaped the cluster.

We thus follow here the second scenario, assuming in addition that the
first stellar generation contains stars with initial masses between
0.1 and 120~M$_{\odot}$, while the second generation consists only of
long-lived low-mass stars (0.1--0.8~M$_{\odot}$). This assumption was
introduced by \citet{Prantzos06} as a simplification and to maximise
the ratio of ejecta to live stars. Simulations of star formation in
the vicinity of rapidly rotating massive stars exhibiting disks are
required to validate this hypothesis on physical grounds. We use a
Salpeter IMF for the first generation of massive stars. On the other
hand, we consider that the present-day log-normal distribution derived
by \citet{Paresce00} reflects the IMF of the long-lived low-mass stars
of both the first and second generation, thus neglecting the probable
scenario that low-mass stars have been preferentially evaporated from
the dynamically evolved cluster \citep[see e.g.][]{DeMarchi00}.

We can then derive the amount of first-generation long-lived stars
that must have been lost in order to fit the high ratio of second to
first generation stars (3:1) found in NGC\,6397. As we consider
pollution by massive star ejecta released on the main sequence only,
we have less polluter material to form new stars than in the case of
NGC\,6752, as studied by \citet{Decressin07b}. This effect is
partially compensated by the higher dilution needed to reproduce the
Li-Na anti-correlation as we form more second generation stars per
unit mass released into slow winds.

Finally, we find that about 90\% of first generation low-mass stars
must have been lost by NGC\,6397, probably during the early event of
gas expulsion by supernovae \citep[see][]{Decressin10}. The initial
mass of NGC\,6397 must thus have been at least ten times higher than
its present day mass. In this estimate, we neglect the long-term
dynamical evolution processes that occur later on, when the two
generations are dynamically mixed, i.e., have the same radial
distribution, and should be ejected from the cluster at the same rate
\citep{Decressin08}. This late evolution can thus not be constrained
by the present fraction of first-to-second generation stars.

\section{Conclusions}
\label{sec:conclusions}
By studying red giant branch stars in the globular cluster NGC\,6397,
we have demonstrated the possibility of distinguishing between the
present stellar generations spectroscopically, by making use of a
double-peaked histogram of Na abundances. A two-population fit returns
a Na abundance similar to the halo field for 25\% of the stars, which
we characterise as belonging to a first generation, whereas the
remaining 75\% (i.e. second generation) have highly elevated Na
abundances. This bimodal abundance signature should be verified for an
extended sample, and similar histograms may also be possible to
produce for N and Al abundances, for which large spreads are present
in globular clusters. The abundance spreads are smaller for O and Mg,
making the task more challenging. Highly precise abundance analysis,
with small statistical error bars will be necessary to identify these
patterns, if indeed present.

Based on 17 different elements heavier than Al, we conclude that there
is no evidence that $\alpha$, iron-peak or neutron-capture abundances
are significantly different between the stellar generations. We have
also estimated the difference in He abundance between two stars of
each generation by enforcing the mass fraction of iron to be same
within the error bars of the analysis. This small (if any) difference
in He is expected from models of cluster self-enrichment and also
supported by the tightness of the main sequence of NGC\,6397. The same
exercise should be performed for other globular clusters, especially
those displaying multiple main sequences, hopefully bringing us closer
to identifying the process responsible for early cluster pollution.

We have finally shown that the abundance patterns observed in
NGC\,6397 can be well reproduced within the ``wind of fast rotating
massive stars" scenario. On the basis of the Li-Na relation we have
been able to infer the dilution factors between the ejecta of massive
polluters and interstellar gas of pristine composition. The observed
ratio of second to first generation stars was used to estimate the
number fraction of first generation long-lived low-mass stars that
must have been lost by NGC\,6397, and thus a lower limit to the
initial mass of this globular cluster.

\acknowledgements{We thank M. Bergemann for helpful advice and
  E. Carretta for communicating data. The referee is thanked for
 many constructive comments to the submitted paper.}

\begin{appendix}
\section{Line list and abundance tables}
\begin{longtable}{llrrl}
      \caption{Line list with reference to the adopted oscillator strength.}\\
\hline\hline
Ion & Wavelength         &   $\chi_l$      & $\log(gf)$  & Ref. \\
    & $[\AA\,]$          &   [eV]          &                &     \\
\hline
\endfirsthead
\caption{Continued.}\\
\hline\hline
Ion & Wavelength         &   $\chi_l$      & $\log(gf)$      & Ref.  \\
    & $[\AA\,]$           &   [eV]          &            &     \\
\hline
\endhead
\hline
\endfoot
O\,I  &   6300.304  & 0.000 & -9.819    &   $^1$           \\ 
O\,I  &   7771.941  & 9.146 &  0.369    &   $^{20}$        \\ 
O\,I  &   7774.161  & 9.146 &  0.223    &   $^{20}$        \\ 
Na\,I &  5682.633  & 2.102 & -0.706    &   $^2$  	 \\ 
Na\,I &  5688.194  & 2.104 & -1.406    &   $^2$		 \\
Na\,I &  5688.205  & 2.104 & -0.452    &   $^2$		 \\
Na\,I &  8183.256  & 2.102 &  0.237    &   $^2$  	 \\ 
Na\,I &  8194.791  & 2.104 & -0.462    &   $^2$		 \\
Na\,I &  8194.824  & 2.104 &  0.492    &   $^2$		 \\
Mg\,I &  5711.088  & 4.346 & -1.833    &   $^3$		 \\
Al\,I &  6696.023  & 3.143 & -1.347    &   $^4$		 \\ 
Al\,I &  6696.185  & 4.022 & -1.576    &   $^5$           \\ 
Al\,I &  6698.673  & 3.143 & -1.647    &   $^4$		 \\
Si\,I &  5665.555  & 4.920 & -1.750    &   $^5$           \\ 
Si\,I &  5690.425  & 4.930 & -1.769    &   $^5$           \\
Si\,I &  6155.134  & 5.619 & -0.754    &   $^5$		 \\
Si\,I &  6237.319  & 5.614 & -0.975    &   $^5$		 \\
Ca\,I &   5349.465  & 2.709 & -0.310   &   $^{6,7}$	 \\ 
Ca\,I &   5512.980  & 2.933 & -0.464   &   $^7$		 \\ 
Ca\,I &   5581.965  & 2.523 & -0.555   &   $^{6,7}$  	 \\
Ca\,I &   5588.749  & 2.526 &  0.358   &   $^{6,7}$   	 \\
Ca\,I &   5590.114  & 2.521 & -0.571   &   $^{6,7}$   	 \\
Ca\,I &   5601.277  & 2.526 & -0.523   &   $^{6,7}$  	 \\
Ca\,I &   5857.451  & 2.933 &  0.240   &   $^7$		 \\
Ca\,I &   6102.723  & 1.879 & -0.793   &   $^8$		 \\ 
Ca\,I &   6122.217  & 1.886 & -0.316   &   $^8$		 \\
Ca\,I &   6162.173  & 1.899 & -0.090   &   $^8$		 \\
Ca\,I &   6166.439  & 2.521 & -1.142   &   $^{6,7}$   	 \\
Ca\,I &   6439.075  & 2.526 &  0.390   &   $^{6,7}$   	 \\
Ca\,I &   6449.808  & 2.521 & -0.502   &   $^{6,7}$   	 \\
Ca\,I &   6471.662  & 2.526 & -0.686   &   $^{6,7}$   	 \\
Ca\,I &   6493.781  & 2.521 & -0.109   &   $^{6,7}$  	 \\
Ca\,I &   6499.650  & 2.523 & -0.818   &   $^{6,7}$  	 \\
Sc\,II&   5239.813  & 1.455 & -0.765   &   $^9$	         \\
Sc\,II&   5526.790  & 1.768 &  0.024   &   $^9$	         \\
Sc\,II&   5641.001  & 1.500 & -1.131   &   $^9$	         \\
Sc\,II&   5669.042  & 1.500 & -1.200   &   $^9$	         \\
Sc\,II&   5684.202  & 1.507 & -1.074   &   $^9$	         \\
Sc\,II&   6245.637  & 1.507 & -1.030   &   $^5$		 \\
Ti\,I &   4981.731  & 0.848 &  0.504   &   $^{10}$        \\
Ti\,I &   4999.503  & 0.826 &  0.250   &   $^{10}$  	 \\
Ti\,I &   5009.645  & 0.021 & -2.259   &   $^{10}$   	 \\
Ti\,I &   5016.161  & 0.848 & -0.574   &   $^{10}$  	 \\
Ti\,I &   5020.026  & 0.836 & -0.414   &   $^{10}$  	 \\
Ti\,I &   5025.570  & 2.041 &  0.250   &   $^{10}$  	 \\
Ti\,I &   5147.478  & 0.000 & -2.012   &   $^{10}$  	 \\
Ti\,I &   5210.385  & 0.048 & -0.884   &   $^{10}$   	 \\
Ti\,I &   5866.451  & 1.067 & -0.840   &   $^{10}$  	 \\
Ti\,I &   6258.102  & 1.443 & -0.355   &   $^{10}$   	 \\
Ti\,I &   6258.706  & 1.460 & -0.240   &   $^{10}$   	 \\
Ti\,II&   4779.985  & 2.048 & -1.260   &   $^5$		 \\
Ti\,II&   4805.085  & 2.061 & -0.960   &   $^5$ 		 \\
Ti\,II&   5005.157  & 1.566 & -2.720   &   $^{11}$   	 \\
Ti\,II&   5013.686  & 1.582 & -2.190   &   $^{11}$   	 \\ 
Ti\,II&   5418.768  & 1.582 & -2.000   &   $^{11}$    	 \\
Cr\,I &   4922.265  & 3.104 &  0.270   &   $^{10}$	 \\
Cr\,I &   5329.138  & 2.914 & -0.064   &   $^{10}$   	 \\
Cr\,II&   5237.329  & 4.073 & -1.160   &   $^{10}$  	 \\
Cr\,II&   5313.563  & 4.074 & -1.650   &   $^{10}$  	 \\
Mn\,I &   4783.427  & 2.298 &  0.042   &   $^{10}$  	 \\
Co\,I &   5369.590  & 1.740 & -1.650   &   $^{12}$	 \\ 
Ni\,I &   4904.407  & 3.542 & -0.170   &   $^{12}$  	 \\
Ni\,I &   4935.831  & 3.941 & -0.350   &   $^{12}$	 \\
Ni\,I &   4953.200  & 3.740 & -0.580   &   $^{13}$	 \\ 
Ni\,I &   4980.166  & 3.606 &  0.070   &   $^{13}$    	 \\
Ni\,I &   5017.568  & 3.539 & -0.020   &   $^{13}$    	 \\
Ni\,I &   5035.357  & 3.635 &  0.290   &   $^{13}$    	 \\
Ni\,I &   5081.107  & 3.847 &  0.300   &   $^{12}$	 \\
Ni\,I &   5082.339  & 3.658 & -0.540   &   $^{12}$	 \\
Ni\,I &   5084.089  & 3.679 &  0.030   &   $^{12}$	 \\
Ni\,I &   5099.927  & 3.679 & -0.100   &   $^{12}$	 \\
Ni\,I &   5115.389  & 3.834 & -0.110   &   $^{12}$	 \\
Ni\,I &   5155.762  & 3.898 &  0.011   &   $^5$		 \\
Ni\,I &   6176.807  & 4.088 & -0.260   &   $^{13}$   	 \\
Cu\,I &   5105.537  & 1.389 & -1.516   &   $^{14}$	 \\ 
Zn\,I &   4810.528  & 4.078 & -0.137   &   $^{15}$	 \\ 
Y\,II &   5087.416  & 1.084 & -0.170   &   $^{16}$	 \\ 
Y\,II &   5200.406  & 0.992 & -0.570   &   $^{16}$	 \\
Zr\,II&   5112.270  & 1.665 & -0.850   &   $^{17}$  	 \\ 
Ba\,II&   5853.668  & 0.604 & -1.000   &   $^{18}$	 \\
Ba\,II&   6496.897  & 0.604 & -0.377   &   $^{18}$   	 \\ 
Ce\,II&   5274.229  & 1.044 & -0.320   &   $^{21}$           \\ 
Nd\,II&   4959.119  & 0.064 & -0.800   &   $^{19}$	 \\ 
Nd\,II&   5092.794  & 0.380 & -0.610   &   $^{19}$	 \\
Nd\,II&   5249.576  & 0.976 &  0.200   &   $^{19}$	 \\
Nd\,II&   5293.163  & 0.823 &  0.100   &   $^{19}$	 \\
Nd\,II&   5319.815  & 0.550 & -0.140   &   $^{19}$        \\ 
Eu\,II&   6645.064  & 1.380 &  0.200   &   $^{22}$        \\    
                   \hline
\multicolumn{2}{l}{$^1$ \citet{Wiese66}}         & \multicolumn{3}{l}{$^{10}$ \citet{Martin88})}     \\
\multicolumn{2}{l}{$^2$ C. Froese Fischer (NIST)}& \multicolumn{3}{l}{$^{11}$ \citet{Pickering01}}  \\
\multicolumn{2}{l}{$^3$ \citet{Lincke71}}        & \multicolumn{3}{l}{$^{12}$ \citet{Fuhr88}}       \\
\multicolumn{2}{l}{$^4$ \citet{Wiese69}}         & \multicolumn{3}{l}{$^{13}$ \citet{Wickliffe97}} \\
\multicolumn{2}{l}{$^5$ Vienna Atomic Line}      & \multicolumn{3}{l}{$^{14}$ \citet{Bielski75}}    \\  
\multicolumn{2}{l}{ \ \ \ \ Database {VALD}}     & \multicolumn{3}{l}{$^{15}$ \citet{Warner68}}     \\
\multicolumn{2}{l}{$^6$ \citet{Smith81}}         & \multicolumn{3}{l}{$^{16}$ \citet{Hannaford82}}  \\ 
\multicolumn{2}{l}{$^7$ \citet{Smith88}}         & \multicolumn{3}{l}{$^{17}$ \citet{Ljung06}}      \\ 
\multicolumn{2}{l}{$^8$ \citet{Smith75}}         & \multicolumn{3}{l}{$^{18}$ \citet{Miles69}}      \\ 
\multicolumn{2}{l}{$^9$ \citet{Lawler89}}        & \multicolumn{3}{l}{$^{19}$ \citet{Komarovskii91}}\\
\multicolumn{2}{l}{$^{21}$ See ref.\,in \citet{Yong05}} & \multicolumn{3}{l}{$^{20}$ \citet{Biemont92}}\\
\multicolumn{2}{l}{$^{22}$ \citet{Lawler01}}\\
\end{longtable}


\begin{sidewaystable}
\tiny
      \caption{Individual star abundances and stellar parameters. The propagated abundance rms error due to uncertainty in equivalent width is written below each table entry. For iron [Fe/H] is given, whereas all other abundances represent [X/Fe].}
      \centering
\begin{tabular}{p{0.7cm}p{0.40cm}rrrrrrrrrrrrrrrrrrrrrrrrr}
\hline\hline
ID    &   $T_{\rm eff}$ & FeII & FeI  & OI &   NaI &   MgI &   AlI &   SiI &   CaI &   ScII &   TiI &   TiII  &  CrI  &  CrII  &  MnI  &  CoI &   NiI &   CuI &   ZnI &   YII &   ZrII &   BaII &   CeII &   NdII & EuII \\
      &  $\log{g}$      &  &   &  & &    &    &    &    &    &    &    &    &    &    &    &    &    &    &    &    &    &  &  & \\   
\hline
12138 &      4727&     -2.06 & -2.10 &  0.45 &   0.14   &  0.26 &  0.51 &  0.30 &  0.26 &  0.13 &  0.19 &  0.31 & -0.14 &  0.01 & -0.43 &  0.01 & -0.17 & -0.76 & -0.02 & -0.26 &  0.09 & -0.07 & 0.39 &  0.17 &  0.49   \\
      &  1.52    &      0.01 &  0.02 &  0.05 &   0.02   &  0.02 &  0.04 &  0.02 &  0.02 &  0.04 &  0.01 &  0.02 &  0.04 &  0.01 &  0.04 &  0.02 &  0.02 &  0.01 &  0.09 &  0.05 &  0.07 &  0.01 & 0.10 &  0.04 &  0.03   \\
16405 &      4790&     -2.06 & -2.19 &  0.41 &   0.29   &  0.22 &  0.57 &  0.23 &  0.22 &  0.18 &  0.15 &  0.28 & -0.18 & -0.04 & -0.49 &  0.01 & -0.19 & -0.81 & -0.01 & -0.26 &  0.01 & -0.06 & 0.40 &  0.18 &  0.39   \\
      &  1.66    &      0.01 &  0.02 &  0.05 &   0.02   &  0.02 &  0.04 &  0.02 &  0.02 &  0.04 &  0.01 &  0.02 &  0.06 &  0.04 &  0.05 &  0.03 &  0.02 &  0.01 &  0.02 &  0.04 &  0.07 &  0.02 & 0.02 &  0.02 &  0.10   \\
14565 &      4811&     -2.09 & -2.17 &  0.47 &   0.22   &  0.26 &  0.51 &  0.22 &  0.20 &  0.12 &  0.17 &  0.31 & -0.14 &  0.01 & -0.50 &  0.09 & -0.17 & -0.82 & -0.03 & -0.21 &  0.12 & -0.03 & 0.48 &  0.21 &  0.42   \\
      &  1.71    &      0.01 &  0.02 &  0.06 &   0.01   &  0.02 &  0.05 &  0.02 &  0.02 &  0.04 &  0.01 &  0.01 &  0.08 &  0.02 &  0.06 &  0.02 &  0.02 &  0.03 &  0.05 &  0.04 &  0.07 &  0.01 & 0.08 &  0.03 &  0.08   \\
5644  &      4826&     -2.06 & -2.12 &  0.66 &  -0.24   &  0.37 &  0.25 &  0.17 &  0.21 &  0.12 &  0.23 &  0.33 & -0.14 &  0.03 & -0.36 &  0.04 & -0.21 & -0.82 & -0.03 & -0.30 &  0.14 & -0.10 & 0.42 &  0.17 &  0.38   \\
      &  1.74    &      0.01 &  0.02 &  0.06 &   0.05   &  0.02 &  9.99 &  0.04 &  0.02 &  0.04 &  0.01 &  0.04 &  0.09 &  0.02 &  0.05 &  0.05 &  0.01 &  0.02 &  0.02 &  0.05 &  0.10 &  0.02 & 0.14 &  0.03 &  0.05   \\
17163 &      4873&     -2.07 & -2.15 &  0.51 &   0.10   &  0.27 &  0.32 &  0.23 &  0.22 &  0.11 &  0.16 &  0.26 & -0.20 &  0.03 & -0.58 & -0.12 & -0.18 & -1.03 & -0.02 & -0.24 &  0.17 & -0.06 &  ... &  0.12 &  0.30   \\
      &  1.84    &      0.02 &  0.02 &  0.13 &   0.04   &  0.03 &  0.10 &  0.04 &  0.03 &  0.04 &  0.03 &  0.03 &  0.17 &  0.04 &  0.14 &  0.12 &  0.04 &  0.09 &  0.03 &  0.04 &  0.08 &  0.03 &  ... &  0.03 &  0.18   \\
20820 &      4905&     -2.10 & -2.15 &  0.53 &   0.14   &  0.30 &  0.47 &  0.27 &  0.27 &  0.12 &  0.19 &  0.30 & -0.09 &  0.00  & -0.40 &  0.11 & -0.16 & -0.91 & -0.04 & -0.27 &  ...  & -0.09 & ... &  0.17 & 0.49    \\
      &  1.92    &      0.01 &  0.02 &  0.08 &   0.03   &  0.02 &  0.05 &  0.02 &  0.02 &  0.04 &  0.01 &  0.04 &  0.09 &  0.03 &  0.04 &  0.05 &  0.02 &  0.05 &  0.01 &  0.06 &  ...  &  0.01 &  ... &  0.06 &  0.05   \\
8952  &      4955&     -2.07 & -2.11 &  0.72 &   0.16   &  0.34 &  0.34 &  0.21 &  0.25 &  0.13 &  0.22 &  0.30 & -0.09 &  0.00 & -0.60 &  0.14 & -0.16 & -0.92 &  0.01 & -0.30 &  ...  & -0.07 &  ... &  0.19 &  0.41   \\
      &  2.04    &      0.02 &  0.02 &  0.12 &   0.04   &  0.04 &  0.10 &  0.04 &  0.03 &  0.04 &  0.02 &  0.05 &  0.09 &  0.04 &  0.12 &  0.05 &  0.03 &  0.04 &  0.03 &  0.06 &  ...  &  0.02 &  ... &  0.03 &  0.11   \\
10737 &      4977&     -2.04 & -2.11 &  0.48 &   0.09   &  0.31 &  0.47 &  0.24 &  0.24 &  0.12 &  0.22 &  0.30 & -0.14 &  0.01 & -0.47 &  0.19 & -0.17 & -0.88 & -0.04 & -0.25 &  ...  & -0.10 &  ... &  0.17 &  0.36   \\
      &  2.10    &      0.02 &  0.02 &  0.08 &   0.04   &  0.04 &  0.09 &  0.04 &  0.02 &  0.04 &  0.03 &  0.04 &  0.07 &  0.04 &  0.07 &  0.07 &  0.03 &  0.07 &  0.03 &  0.06 &  ...  &  0.02 &  ... &  0.04 &  0.12   \\
13006 &      5002&     -2.03 & -2.09 &  0.54 &   0.04   &  0.31 &  0.14 &  0.19 &  0.25 &  0.10 &  0.20 &  0.27 & -0.11 & -0.01 & -0.43 &  0.16 & -0.17 & -0.80 & -0.03 & -0.28 &  ...  & -0.12 &  ... &  0.11 &   ...   \\
      &  2.16    &      0.02 &  0.02 &  0.09 &   0.04   &  0.04 &  0.17 &  0.04 &  0.02 &  0.04 &  0.03 &  0.04 &  0.11 &  0.05 &  0.05 &  0.08 &  0.04 &  0.04 &  0.04 &  0.02 &  ...  &  0.03 &  ... &  0.04 &   ...   \\
17691 &      5003&     -2.01 & -2.05 &  0.78 &  -0.25   &  0.49 &  0.30 &  0.19 &  0.29 &  0.11 &  0.26 &  0.32 & -0.04 &  0.06 & -0.42 &  0.20 & -0.14 & -0.92 &  0.07 & -0.27 &  ...  & -0.20 &  ... &  0.04 &   ...   \\
      &  2.16    &      0.02 &  0.02 &  0.11 &   0.06   &  0.04 &  9.99 &  0.02 &  0.03 &  0.04 &  0.02 &  0.04 &  0.10 &  0.04 &  0.05 &  0.05 &  0.03 &  0.11 &  0.04 &  0.05 &  ...  &  0.03 &  ... &  0.03 &   ...   \\
21284 &      5044&     -2.10 & -2.15 &  0.57 &   0.11   &  0.34 &  0.45 &  0.17 &  0.25 &  0.12 &  0.22 &  0.31 &  0.01 &  0.03 & -0.64 &  0.26 & -0.13 & -0.98 &  0.02 & -0.25 &  ...  & -0.10 &  ... &  0.21 &   ...   \\
      &  2.27    &      0.01 &  0.02 &  0.12 &   0.04   &  0.04 &  0.12 &  0.06 &  0.01 &  0.04 &  0.01 &  0.04 &  0.06 &  0.05 &  0.11 &  0.07 &  0.01 &  0.08 &  0.03 &  0.04 &  ...  &  0.02 &  ... &  0.05 &   ...   \\
9424  &      5047&     -2.08 & -2.18 &  0.67 &   0.15   &  0.28 &  0.43 &  0.30 &  0.23 &  0.09 &  0.18 &  0.26 & -0.06 &  0.03 & -0.57 &  0.10 & -0.17 & -0.92 & -0.04 & -0.29 &  ...  & -0.11 &  ... &  0.21 &   ...   \\
      &  2.30    &      0.02 &  0.02 &  0.19 &   0.04   &  0.04 &  0.12 &  0.06 &  0.02 &  0.04 &  0.03 &  0.06 &  0.04 &  0.06 &  0.13 &  0.06 &  0.03 &  0.14 &  0.04 &  0.02 &  ...  &  0.04 &  ... &  0.05 &   ...   \\
16248 &      5049&     -2.11 & -2.17 &  0.38 &   0.14   &  0.21 &  0.45 &  0.31 &  0.25 &  0.13 &  0.20 &  0.31 & -0.08 &  0.05 & -0.49 &  0.06 & -0.12 & -1.04 &  0.04 & -0.23 &  ...  & -0.10 &  ... &  0.25 &   ...   \\
      &  2.31    &      0.03 &  0.02 &  0.14 &   0.05   &  0.04 &  0.09 &  0.03 &  0.03 &  0.06 &  0.04 &  0.04 &  0.09 &  0.06 &  0.20 &  0.14 &  0.04 &  0.08 &  0.03 &  0.03 &  ...  &  0.03 &  ... &  0.10 &   ...   \\
11363 &      5144&     -2.11 & -2.14 &  0.67 &   0.19   &  0.36 &  0.26 &  0.23 &  0.27 &  0.12 &  0.25 &  0.33 & -0.08 &  0.07 & -0.92 &  ...  & -0.11 &  ...  & -0.11 & -0.25 &  ...  & -0.10 &  ... &  0.22 &   ...   \\
      &  2.49    &      0.02 &  0.02 &  0.27 &   0.05   &  0.04 &  0.22 &  0.06 &  0.02 &  0.04 &  0.03 &  0.03 &  0.09 &  0.09 &  0.22 &  ...  &  0.03 &  ...  &  0.12 &  0.04 &  ...  &  0.03 &  ... &  0.11 &   ...   \\
9649  &      5166&     -2.10 & -2.13 &  0.73 &   0.22   &  0.30 &  0.38 &  0.25 &  0.27 &  0.11 &  0.26 &  0.29 & -0.07 &  0.05 & -0.56 &  ...  & -0.14 &  ...  &  0.02 & -0.29 &  ...  & -0.09 &  ... &  0.23 &   ...   \\
      &  2.52    &      0.02 &  0.02 &  9.99 &   0.05   &  0.04 &  0.18 &  0.06 &  0.02 &  0.06 &  0.03 &  0.05 &  0.04 &  0.07 &  0.04 &  ...  &  0.04 &  ...  &  0.04 &  0.02 &  ...  &  0.04 &  ... &  0.09 &   ...   \\
14592 &      5175&     -2.09 & -2.13 &  0.69 &  -0.04   &  0.43 &  0.47 &  0.25 &  0.25 &  0.08 &  0.29 &  0.29 & -0.08 &  0.01 & -0.65 &  ...  & -0.10 &  ...  &  0.03 & -0.30 &  ...  & -0.06 &  ... &  0.20 &   ...   \\
      &  2.59    &      0.03 &  0.02 &  0.19 &   0.10   &  0.04 &  9.99 &  0.04 &  0.03 &  0.08 &  0.04 &  0.09 &  0.08 &  0.09 &  0.08 &  ...  &  0.04 &  ...  &  0.06 &  0.05 &  ...  &  0.03 &  ... &  0.07 &   ...   \\
4680  &      5180&     -2.12 & -2.15 &  0.77 &   0.21   &  0.36 &  0.39 &  0.17 &  0.29 &  0.12 &  0.29 &  0.31 &  0.03 &  0.05 & -0.60 &  ...  & -0.10 &  ...  &  0.11 & -0.20 &  ...  & -0.08 &  ... &  ...  &   ...   \\
      &  2.60    &      0.02 &  0.02 &  0.17 &   0.06   &  0.05 &  9.99 &  0.05 &  0.02 &  0.05 &  0.02 &  0.04 &  0.09 &  0.08 &  0.05 &  ...  &  0.04 &  ...  &  0.05 &  0.03 &  ...  &  0.04 &  ... &  ...  &   ...   \\
12294 &      5258&     -2.09 & -2.11 &  ...  &   0.24   &  0.29 &  0.71 &  0.22 &  0.24 &  0.13 &  0.28 &  0.32 &  0.12 &  0.12 &  ...  &  ...  & -0.16 &  ...  &  0.16 & -0.23 &  ...  & -0.06 &  ... &  ...  &   ...   \\
      &  2.87    &      0.04 &  0.02 &  ...  &   0.08   &  0.08 &  9.99 &  0.12 &  0.04 &  0.06 &  0.05 &  0.04 &  0.16 &  0.10 &  ...  &  ...  &  0.06 &  ...  &  0.08 &  0.11 &  ...  &  0.04 &  ... &  ...  &   ...   \\
7658  &      5273&     -2.06 & -2.12 &  ...  &   0.28   &  0.23 &  0.57 &  0.39 &  0.25 &  0.07 &  0.22 &  0.25 & -0.14 &  0.04 &  ...  &  ...  & -0.09 &  ...  &  0.01 & -0.27 &  ...  & -0.10 &  ... &  ...  &   ...   \\
      &  2.91    &      0.03 &  0.02 &  ...  &   0.06   &  0.09 &  9.99 &  0.08 &  0.04 &  0.06 &  0.05 &  0.06 &  0.09 &  0.20 &  ...  &  ...  &  0.04 &  ...  &  0.05 &  0.13 &  ...  &  0.06 &  ... &  ...  &   ...   \\
11142 &      5279&     -2.09 & -2.11 &  ...  &   0.17   &  0.33 &  0.42 &  0.12 &  0.24 &  0.14 &  0.25 &  0.27 &  0.07 &  0.06 &  ...  &  ...  & -0.13 &  ...  &  0.03 & -0.22 &  ...  & -0.06 &  ... &  ...  &   ...   \\
      &  2.93    &      0.02 &  0.02 &  ...  &   0.07   &  0.08 &  0.23 &  0.18 &  0.02 &  0.04 &  0.04 &  0.04 &  0.07 &  0.12 &  ...  &  ...  &  0.04 &  ...  &  0.06 &  0.02 &  ...  &  0.03 &  ... &  ...  &   ...   \\
7728  &      5288&     -2.09 & -2.11 &  ...  &   0.04   &  0.28 &  0.61 &  0.40 &  0.24 &  0.05 &  0.27 &  0.34 &  0.11 & -0.11 &  ...  &  ...  & -0.07 &  ...  &  0.13 & -0.23 &  ...  &  0.06 &  ... &  ...  &   ...   \\
      &  2.95    &      0.04 &  0.02 &  ...  &   0.10   &  0.08 &  0.17 &  0.08 &  0.04 &  0.04 &  0.05 &  0.05 &  0.07 &  0.26 &  ...  &  ...  &  0.06 &  ...  &  0.06 &  0.09 &  ...  &  0.06 &  ... &  ...  &   ...   \\
\hline
\end{tabular}
\end{sidewaystable}

\end{appendix}

\bibliographystyle{aa}

\end{document}